\documentclass[aps,prb,superscriptaddress,twocolumn,showpacs]{revtex4-1}
\usepackage{graphicx}
\usepackage{tikz}

\usetikzlibrary{
arrows,shapes.misc,shapes.arrows,matrix,
positioning,scopes,decorations.pathmorphing,shadows,
shapes.geometric,decorations.pathreplacing,shapes.gates.logic.US,calc}

\tikzset{
matStyle/.style={column sep=.4mm,row sep=.4mm,font=\itshape\huge,execute at empty cell=\node{\phantom{.}};},
measStyle/.style={and gate US,text centered,minimum size=8mm,thick,draw=black,
    top color=white,bottom color=green!10!white},
meterStyle/.style={rectangle,text centered,minimum size=8mm,thick,draw=black,
    top color=white,bottom color=green!10!white},
lblStyle/.style={rectangle,minimum size=8mm,thick,draw=white,
	top color=white,bottom color=white},
rawStyle/.style={},
prepStyle/.style={rectangle,minimum size=8mm,thick,draw=white,text=red!60!black,
	top color=white,bottom color=white},
ctrlStyle/.style={circle,text centered,inner sep=-1pt,minimum size=1mm,thick,draw=black,
	top color=black,bottom color=black},
ctrloStyle/.style={circle,text centered,inner sep=-1pt,minimum size=1mm,thick,draw=black,
	top color=white,bottom color=white},
targStyle/.style={circle,text centered,inner sep=0pt,minimum size=3mm,thick,draw=black,
	top color=white,bottom color=white,append after command={
        [shorten >=\pgflinewidth, shorten <=\pgflinewidth, thick,]
        (\tikzlastnode.north) edge[thick] (\tikzlastnode.south)
        (\tikzlastnode.east) edge[thick] (\tikzlastnode.west)
        }},
qswapStyle/.style={circle,text centered,inner sep=0pt,minimum size=3mm,thin,draw=white,opacity=0,
	top color=white,bottom color=white,append after command={
        [shorten >=\pgflinewidth, shorten <=\pgflinewidth, thick,color=black]
        (\tikzlastnode.north east) edge[thick] (\tikzlastnode.south west)
        (\tikzlastnode.north west) edge[thick] (\tikzlastnode.south east)
        }},
gateStyle/.style={rectangle,minimum size=8mm,thick,draw=black,
	top color=white,bottom color=blue!10!white},
mgStyle/.style={rectangle,minimum size=8mm,transparent},
mgBoxStyle/.style={rectangle,minimum size=8mm,thick,draw=black,
	top color=white,bottom color=blue!10!white},
qwStyle/.style={thick,draw=black,font=\ttfamily},
qwxStyle/.style={thick,draw=black,font=\ttfamily},
>=latex,thick,
/pgf/every decoration/.style={/tikz/sharp corners},
pointStyle/.style={coordinate},
>=stealth',thick,draw=black,tip/.style={->,shorten >=0.007pt},
every join/.style={rounded corners},
hv path/.style={to path={-| (\tikztotarget)}},
vh path/.style={to path={|- (\tikztotarget)}},
text height=1.5ex,text depth=.25ex
}

\pgfdeclarelayer{background}		
\pgfdeclarelayer{foreground}		
\pgfsetlayers{background,main,foreground}

\newcommand{\gate}[2][] {\node (#1) [gateStyle] {\phantom{$#2$}} node [yshift=-5pt] {$#2$};}
\newcommand{\lbl}[2][]{\node (#1) [lblStyle] {$#2$};}
\newcommand{\raw}[2][]{\node (#1) [rawStyle] {$#2$};}
\newcommand{\prep}[2][]{\node (#1) [prepStyle] {$#2$};}
\newcommand{\mg}[2][]{\node (#1) [mgStyle] {\phantom{$#2$}};}
\newcommand{\point}[1][]{ \node (#1) [pointStyle] {};}
\newcommand{\targ}[1][]{ \node (#1) [targStyle] {};}
\newcommand{\qswap}[1][]{ \node (#1) [qswapStyle] {};}
\newcommand{\ctrl}[1][]{ \node (#1) [ctrlStyle] {};}
\newcommand{\ctrlo}[1][]{ \node (#1) [ctrloStyle] {};}
\newcommand{\meas}[2][]{\node (#1) [measStyle] {$#2$};}

\newcommand{\meter}[1][] {
	\node (#1) [meterStyle] {\phantom{$X$}};
	\draw[thick] ++(.25,-.2) arc (0:180:.25 and .1);
	\draw[thick] (0,-.14) -- +(.1,.2)
	;}

\newcommand{\multigate}[4][]{
    \draw[gateStyle,name=#1] (#2.north west) rectangle (#3.south east);
    \draw[draw=black,font=\itshape\LARGE] ($0.5*(#2)+0.5*(#3)$) node {$#4$};
}

\newcommand{\qw}[2]{\draw[qwStyle] (#1) -- (#2);}
\newcommand{\cw}[2]{\draw[qwStyle, double] (#1) -- (#2);}

\newcommand{\qwx}[2]{\draw[qwxStyle] (#1) -- (#2);}
\newcommand{\cwx}[2]{\draw[qwxStyle, double] (#1) -| (#2);}
\newcommand{\dwx}[2]{\draw[qwxStyle, dotted] (#1) -- (#2);}
\newcommand{\fwx}[2]{\draw[qwxStyle] (#1) -- (#2) node[midway,left] {$\mathcal{F}$};}

 \newcommand{\qwxB}[2]{\draw[qwxStyle] (#1)  to [bend left=4]  (#2);}
 \newcommand{\dwxB}[2]{\draw[qwxStyle,dotted] (#1)  to [bend left=4]  (#2);}
 \newcommand{\cwxB}[2]{\draw[qwxStyle,double] (#1)  to [bend left=4]  (#2);}

\newcommand{\bra}[1]{{\left\langle{#1}\right\vert}}
\newcommand{\ket}[1]{{\left\vert{#1}\right\rangle}}
\usepackage{amsmath, amsthm, amssymb}
\usepackage{graphicx}
\usepackage{color}
\usepackage{multirow}

\usepackage{natbib}
\usepackage{hyperref}
\hypersetup{
        colorlinks=true,
}

\newcommand{\be}{\begin{equation}}
\newcommand{\ee}{\end{equation}}

\begin{document}

\title{Solving strongly correlated electron models on a quantum computer}

\author{Dave Wecker}
\affiliation{Quantum Architectures and Computation Group, Microsoft Research, Redmond, WA 98052, USA}

\author{Matthew~B.~Hastings}
\affiliation{Station Q, Microsoft Research, Santa Barbara, CA 93106-6105, USA}
\affiliation{Quantum Architectures and Computation Group, Microsoft Research, Redmond, WA 98052, USA}

\author{Nathan Wiebe}
\affiliation{Quantum Architectures and Computation Group, Microsoft Research, Redmond, WA 98052, USA}

\author{Bryan K. Clark}
\affiliation{Station Q, Microsoft Research, Santa Barbara, CA 93106-6105, USA}
\affiliation{UIUC}

\author{Chetan Nayak}
\affiliation{Station Q, Microsoft Research, Santa Barbara, CA 93106-6105, USA}

\author{Matthias Troyer}
\affiliation{Theoretische Physik, ETH Zurich, 8093 Zurich, Switzerland}


\begin{abstract}
One of the main applications of future quantum computers will be the simulation of quantum models. While the evolution of a quantum state under a Hamiltonian is straightforward (if sometimes expensive), using quantum computers to determine the ground state phase diagram of a quantum model and the properties of its phases is more involved. Using the Hubbard model as a prototypical example, we here show all the steps necessary to determine its phase diagram and ground state properties on a quantum computer. In particular, we discuss strategies for efficiently determining and preparing the ground state of the Hubbard model starting from various mean-field states with broken symmetry. We present an efficient procedure to prepare arbitrary Slater determinants as initial states and present the complete set of quantum circuits needed to evolve from these to the ground state of the Hubbard model. We show that, using efficient nesting of the various terms each time step in the evolution can be performed with just $\mathcal{O}(N)$ gates and $\mathcal{O}(\log N)$ circuit depth. We   give explicit circuits to measure arbitrary local observables and static and dynamic correlation functions, both in the time and frequency domain.  We further present efficient non-destructive approaches to measurement that avoid the need to re-prepare the ground state after each measurement and that quadratically reduce the measurement error.
\end{abstract}

\maketitle

\section{Introduction}
Feynman envisioned that quantum computers would enable the simulation of quantum systems: an initial
quantum state can be unitarily evolved with a quantum computer using resources that are polynomial in the size of the system and the
evolution time.\cite{Feynman1982} However, it is not clear that this would enable us to determine the answers to
the questions of greatest interest to condensed matter physicists. The properties most readily measured in
experiments do not appear to be simple to determine in a quantum simulation, and known algorithms
for quantum simulation do not return the quantities of greatest physical interest.
In experiments in a solid, one can rarely, if ever, prepare a known
simple initial state, nor does one know precisely the Hamiltonian with which it would evolve. Meanwhile,
the ground state energy (or the energy of any eigenstate yielded by quantum phase estimation \cite{Kitaev95,Kitaev02,Nielsen10}) is
not a particularly enlightening quantity and, furthermore, given a quantum state it is not clear how to
determine its long-wavelength `universal' properties.
Consider, for instance, the following questions: given a model Hamiltonian for the electrons in a solid, such as the Hubbard model,
can we determine if its ground state is superconducting? Can we determine why it is superconducting?
In this paper, we show that it is possible, in principle, to answer the first question and, if it has a clear answer (in a sense to be discussed),
the second question as well. Moreover, we show that it is not only possible in principle but, in fact, feasible to answer such questions
with a quantum computer of moderate size.

The first step in the solution of a quantum Hamiltonian is to map its Hilbert space to the states
of a quantum computer.
While the Hilbert spaces of systems of bosons or spins are mapped most naturally to the states of a general-purpose
quantum computer, fermionic systems can also be simulated by using a Jordan-Wigner transformation to represent them
as spin systems\cite{Ortiz2001,whitfield2011}; the cost can be reduced to $\log(N)$ using additional qubits\cite{Bravyi02} or using appropriate ordering techniques discussed below.
In order to evolve the system
on a general purpose quantum computer, we decompose the time-evolution operator according to the Trotter-Suzuki decomposition
\cite{Trotter59,Suzuki76}; the number of time intervals in such a decomposition is determined by the desired accuracy.
The evolution for each time interval is expressed in terms of the available gates. If the Hamiltonian
can be broken into $k$ non-commuting terms, then the gates can be broken into $k$ sets; within each set, the gates can be applied in parallel,
and only ${\mathcal O}(k)$ time steps are needed for each interval. Time evolution can be used in conjunction with the quantum phase estimation
\cite{Kitaev95,Kitaev02,Nielsen10} algorithm to find an approximate energy eigenvalue and eigenstate of the Hamiltonian.

The challenge when applying this approach to electronic systems is that the Coulomb Hamiltonian in second quantized form
\begin{equation}
\label{eq:electronic}
H=\sum_{ p,q=1}^Nh_{pq}c^\dag_{p}c_{q}  + \frac{1}{2}\sum_{p,q,r,s=1}^Nh_{pqrs}c^\dag_{p}c^\dag_{q}c_{r}c_{s}.
\end{equation}
has $k={\mathcal O}(N^4)$ terms for $N$ orbitals. Here the operators $c^\dag_{p}$ and $c_{p}$ create and annihiliate an electron with spin $\sigma$ in spin-orbital $p$ (combining orbital index $i$ and spin index $\sigma$). The quadratic $h_{pq}$-terms  arise from the kinetic energy of the electrons and the Coulomb potential due to the nuclei (which are assumed to be classical in the usual Born-Oppenheimer approximation) and the quartic $h_{pqrs}$-terms encode the Coulomb repulsion. In order to estimate the ground state energy of a system of
interacting electrons in a large molecule of $N$ orbitals, naive estimates indicated that
$\mathcal{O}(N^{8})$ operations are needed \cite{Wecker14a}, due to the  large number of non-commuting terms in the Hamiltonian, but
more recent analyses indicate that $\sim N^5$ operations may be sufficient.\cite{Hastings15,Poulin14} While this makes the simulation of small classically-intractable molecules feasible, the scaling is still too  demanding to perform electronic structure calculations for more than a few hundred orbitals. This makes the brute-force simulation of large molecules and crystalline solids impractical.

One approach to reduce the scaling complexity for the simulation of crystalline  solids is to focus on effective models that capture only the relevant bands close to the Fermi energy, and simplify the ${\mathcal O}(N^4)$ terms of the Coulomb interaction to just a few dominant terms. One of the paradigmatic effective models for strongly correlated fermions, discussed in detail in Sec. \ref{sec:hubbard}, is the square lattice Hubbard model, which is a radical simplification of the full electronic structure Hamiltonian (\ref{eq:electronic}) to arguably the simplest interacting  single-band model, described by the Hamiltonian
\begin{eqnarray}
H_{\rm Hub} &=&-\sum_{\langle i,j \rangle}\sum_\sigma t_{ij}\left(c^\dag_{i,\sigma}c_{j,\sigma} + c^\dag_{j,\sigma}c_{i,\sigma} \right) \nonumber \\
&&+ U \sum_i n_{i,\uparrow}n_{i,\downarrow} + \sum_i \epsilon_i n_i,
\label{eq:hubbard}
\end{eqnarray}
where $n_{i,\sigma}=c^\dag_{i,\sigma}c_{i,\sigma}$ is the local spin density and $n_i=\sum_\sigma n_{i,\sigma}$ is the total local density. Of all the orbitals in a unit cell, we only keep a {\it single} orbital (describing {\it e.g.} the Cu $d_{x^2-y^2}$ in a cuprate superconductor), reduce the long-range Coulomb repulsion to just the local repulsion $U$ within this orbital, and the hybridizations $t_{ij}=-h_{pq}$ to nearest neighbors on the lattice. The on-site energies will be called $\epsilon_i=h_{ii}$. Usually one can consider a translationally-invariant model and then drop the indices in $t_{ij}$ and $\epsilon_i$ since all sites are equivalent. Here we will keep them explicitly since
they may not all be the same during the adiabatic evolution.

The computational complexity of the Hubbard model (\ref{eq:hubbard}) is much reduced compared to the full electronic structure problem (\ref{eq:electronic}), since only a single orbital is used per unit cell instead of dozens needed for the full problem and the number of terms is linear in $N$ instead of scaling with the fourth power $N^4$. On a lattice of $20\times20$ unit cells this means a reduction of the total number of terms from about $\sim10^{12}$ to $\sim 10^3$, which makes such a simulation feasible on a quantum computer. Furthermore, since the 2D Hubbard Hamiltonian can be expressed as a sum of five non-commuting terms
(the on-site interaction terms and four sets of hopping terms,
one for each nearest neighbor to a given site) each time step in the Trotter-Suzuki decomposition
requires only  $\sim \log N$ parallel circuit depth (in fact, constant, if it were not for the Jordan-Wigner strings). To achieve this optimal scaling we use a Jordan-Wigner transform combined with optimal term ordering and the ``nesting" and Jordan-Wigner cancellation technique of Ref.~\onlinecite{Hastings15}.

The simplifying features of the Hubbard model that afford such advantageous scaling with $N$ are
the restriction to nearest-neighbor hopping and on-site interactions. Of course, a real solid will have long-ranged Coulomb interactions
and longer-ranged hopping terms. However, the point of a model such as the Hubbard model is not to give a quantitatively accurate
description of a solid, but rather to capture some essential features of strongly-correlated
electrons in transition-metal oxides. For these purposes, a simplified model that does not
include all of the complexity of a real solid should be sufficient (although, for the case of the cuprate high-$T_c$ superconductors,
the required simplified model may be a bit more complicated than the Hubbard model; we focus here on the Hubbard model for
illustrative purposes). 

A central question in the study of a model of strongly correlated electrons is: what is its phase diagram as a function of the parameters in
the model? The Hubbard model has the following interesting subquestion: is there a superconducting phase somewhere in the phase diagram? 
Once the phase diagram has been determined, we wish to characterize the phases in it by computing key quantities such as the quasiparticle energy gap
and phase stiffness (superfluid density) in a superconducting phase. To answer these questions we will need to determine the ground state wave function for a range of interaction parameters and densities and then  measure  ground state correlation functions.

 We take the following approach to solve the Hubbard model (or related models) on a quantum computer: 
 
 \begin{enumerate}
 \item Adiabatically prepare an approximate ground state $|\tilde\Psi_0\rangle$ of the Hubbard model starting from different initial states.
\item Perform a quantum phase estimation to project the true ground state wave function $|\Psi_0\rangle$ from the approximate state $|\tilde\Psi_0\rangle$, and measure the ground state energy $E_0=\langle \Psi_0|H|\Psi_0\rangle$.
\item Measure local observables and static and dynamic correlation functions of interest
\end{enumerate}

The general strategy for simulating quantum models has already been discussed previously, starting from Feynman's  original proposal.\cite{Feynman1982} Abrams and Lloyd \cite{Abrams:1997ha} were the first to discuss time evolution under the Hubbard Hamiltonian, but without giving explicit quantum circuits; in Ref. \onlinecite{Abrams:1999jv} the same authors discuss how quantum phase estimation can be used to project from a trial state into an energy eigenstate. Refs. \onlinecite{Ortiz2001} and \onlinecite{Somma2002} provided more details of how to simulate quantum lattice models, proposed an algorithm  to construct arbitrary Slater determinants and discussed measurements of various observables of interest.  The idea of adiabatic state preparation for interacting fermion systems was proposed in Ref.~\onlinecite{Ortiz2001} and adiabatic preparation has been considered by a number of authors\cite{Farhi,Trebst2005,HL}

In this paper we go beyond the earlier work in several crucial aspects. Besides presenting full details of the necessary quantum circuits for time evolution (Sec.~\ref{sec:timeevolve}) and measurements (Secs. \ref{sec:measurements} and \ref{sec:non-destructive}) we address crucial issues that have not received much attention so far. In particular, in Sec. \ref{sec:prepare} we discuss in detail how to adiabatically prepare approximate ground states $|\tilde\Psi_0\rangle$ of the Hubbard model starting from different initial states, capturing various proposed broken symmetries, and use this to determine the phase diagram of the Hubbard model. Having thereby deduced the correct initial state, we can cheaply prepare multiple copies of the ground state of the system. 

Due to the benign scaling of circuit depth with $\log N$, the ground state of the Hubbard model on large lattices can be prepared in very short time of less than a second, even assuming logical gate times in the $\mu s$ range as we discuss in Sec. \ref{sec:qpe}.

We then present, in Sec. \ref{sec:measurements}, details of how to measure various quantities of interest, including both equal time and dynamic correlation functions. For the latter, we present a new approach, based on importance sampling in the frequency domain to complement the usual approach of measuring in the time domain.

Since each measurement only gives limited information, the ground state has to be prepared many times, which leads to a significant total runtime of a hypothetical quantum computer. In Sec. \ref{sec:non-destructive}, we present
new approaches to non-destructive measurements of arbitrary observables that substantially reduce the runtime.
By using a relatively cheap quantum phase estimation to recreate the ground state after a measurement, we
reduce the number of times the ground state needs to be prepared from scratch. Furthermore their runtime scales as $\mathcal{O}(\epsilon^{-1})$ with the desired accuracy $\epsilon$ instead of the  $\mathcal{O}(\epsilon^{-2})$ scaling of the na\"ive approach to measurements. 

For the aficionado of quantum algorithms, the paper contains the following new algorithmic ideas. In Sec. \ref{sec:qpe}, we introduce a variant of quantum phase estimation that gives a two-fold reduction in the
number of time steps and a four-fold reduction in rotation gates compared to the standard approach. Section \ref{sec:prepare} presents an approach to speeding up adiabatic state preparation by adiabatically changing the Trotter time step. In Sec.  \ref{sec:measurements}, we present an algorithm for measuring dynamical correlation functions by importance sampling of the spectral function in the frequency domain. Finally, Sec. \ref{sec:non-destructive} presents two approaches to non-destructive measurements.

\section{The Hubbard model}

\label{sec:hubbard}
\subsection{Derivation of the model}

The square lattice Hubbard model is one of the paradigmatic models for strongly interacting fermions.
Although it was originally introduced as a model for
itinerant-electron ferromagnetism \cite{Gutzwiller63,Hubbard63}, it is now studied
primarily as a model for the Mott transition and broken-symmetry (and, perhaps, even topological) ordering
phenomena in the vicinity of this transition (for recent reviews, see, for instance,  Refs. \onlinecite{Sachdev10,Fradkin14}).
In particular, it has been hypothesized that the square lattice Hubbard model
captures the important ingredients of the cuprate high temperature superconductors.
It is a radical simplification of the full Hamiltonian of a solid. A complete description of
a solid takes the form:
\begin{multline}
H = \sum_i \frac{p_i^2}{2 m_e} + \sum_a \frac{P_a^2}{2 M_a} + 
\sum_{i>j} \frac{e^2}{|{r_i}-{r_j}|} \\+ \sum_{i,a} \frac{Z_a e^2}{|{r_i}-{R_a}|} 
+ \sum_{a>b} \frac{Z_a Z_b e^2}{|{R_a}-{R_b}|}
\end{multline}
where $i, j = 1, \ldots, N_e$ and $a, b =1, \ldots N_\text{ions}$; $N_e$ and $N_\text{ions}$ are, respectively,
the number of electrons and ions; and $Z_a$ is the atomic number of ion $a$ and $M_a$ its mass.
In certain situations, the physics is dominated by electron-electron interactions.
The ions form a crystalline lattice and the vibrations of this lattice (phonons), though quantitatively
important, do not, it is hypothesized, play a major qualitative role.
Then we may, in a first approximation, ignore the dynamics of the ions and, thereby, obtain
a purely electronic Hamiltonian which, in general, will take the form of Eq. (\ref{eq:electronic}). Explicitly introducing labels for spin, unit cells and orbitals with a unit cell of a periodic solid we write this Hamiltonian as
\begin{multline}
H =-\!\!\sum_{i,j ; a,b; \sigma} t^{ab}_{ij}\left(c^\dag_{i,a,\sigma}c^{}_{j,b,\sigma}
+ \text{h.c.} \right)  + \sum_{i,a} \epsilon_{a} n_{i,a},
\\
+  \sum_{i,a} U_a n_{i,a} (n_{i,a} - 1)/2\\
+ \sideset{}{'}\sum_{i,j,k,l;a,b,c,d}^{} V_{ijkl}^{abcd} c^\dagger_{i,a} c^\dagger_{j,b} c_{k,c}  c_{l,d} +\ldots 
\label{eq:general-lattice-model}
\end{multline}
Here, $i,j$ label unit cells and $a,b$ label orbitals within a unit cell.
The
local density
in orbital $a$ is $n_{i,a}=\sum_\sigma n_{i,a,\sigma}$.
The coupling constants have the following meanings:
$t_{i,j}^{a,b}$ is the matrix
element for an electron to hop from orbital $b$ in unit cell $j$ to orbital $a$ in unit cell $i$ (possibly
two different orbitals within the same unit cell);
$\epsilon_{a}$ is the energy of orbital $a$ in isolation; 
$U_a$ is the energy penalty for two electrons to occupy the same orbital in the same unit cell.
This term can be rewritten equivalently as $\sum_{i,a} U_a n_{i,a,\uparrow} n_{i,a,\downarrow}$.
The prime on the summation in the final line indicates that we have omitted the term with $i=j=k=l$ and $a=b=c=d$,
which has been separately included as the $U_a$ term.
They include, for instance, the terms with
$i=k \neq j=l$, $a=c \neq b=d$, which is the density-density interaction $V_{ijij}^{abab} n_{i,a} n_{j,b}$,
and the terms with $j=l$, $b=d$, which is correlated hopping term
$V_{ijkj}^{abcb} c^\dagger_{i,a} c_{k,c} n_{j,b}$. These terms will be assumed to be much
smaller than $t_{i,j}^{a,b}$, $U_a$ and will be dropped. 

If the differences between the $\epsilon_a$s are the largest energy scales in
the problem, then there will be, at most, a single partially-filled orbital and we can ignore all of the
other orbitals when studying the physics of the model at temperatures less than $|{\epsilon_a} - {\epsilon_b}|$. These energy differences
can be on the order of eV and, therefore, such an approximation will be valid over a very large range of
temperatures and energies. Similarly, if $t^{ab}_{ii}$ is a large energy scale, it may be possible to reduce the model
to one with a single orbital per unit cell (which, in this case, will be a linear combination of the orbitals
in Eq. (\ref{eq:general-lattice-model})). If the on-site interactions $U$ in this orbital (we drop the subscript
$a$ since there is only a single orbital under consideration) are much larger than the interactions with nearby unit cells
(nearest-neighbor, next-nearest neighbor, etc.), then we can drop the latter. Similarly, if the nearest-neighbor
hopping is much larger than more distant hopping matrix elements, then we can drop the latter and write down the simplified model of Eq. (\ref{eq:hubbard}).
This is the Hubbard model.

In the case of the copper-oxide superconductors, the only orbital that is retained
is the Cu $d_{x^2-y^2}$ orbital. (However, it has also been argued that a $3$-band model including
O $p_x$ and $p_y$ orbitals is necessary to capture the physics of the cuprates.\cite{Varma97})
Usually one can drop the indices in $t_{ij}$, $U_i$ and $\epsilon_i$ since all sites are equivalent,
but we will keep them here since during the adiabatic solution they will not
all be the same.

Due to the presumed separation of energy scales in (\ref{eq:general-lattice-model}),
the effective model (\ref{eq:hubbard}) is much simpler than 
(\ref{eq:general-lattice-model}) and has a much smaller Hilbert space.
As we will discuss in detail in this paper, this simplification makes simulation of $20\times20$ or more unit cells
feasible on a quantum computer.


\subsection{The physics of the Hubbard model}

At half-filling (one electron per unit cell) the Hubbard model gives a simple account of Mott insulating behavior.
\footnote{A system in which insulating behavior is caused by electron-electron interactions in a half-filled band
(which would otherwise be metallic) is called a Mott insulator.}
For $U=0$, the ground state is metallic. There is a single band and it is half-filled; the Fermi surface is the diamond $|{k_x}\pm {k_y}|=\pi/a$,
where $a$ is the lattice constant. However, for $U>0$, the ground state is insulating.
While this is true for all $U>0$, the physics is different in the small $U \ll t$ and $U\gg t$ limits.
For $U \ll t$, the system lowers its energy by developing antiferromagnetic order, as a result of which the unit cell doubles in size. There are now
two electrons per unit cell and the system is effectively a band insulator. The antiferromagnetic moment $N$ and the charge gap $\Delta_\text{charge}$ are both
exponentially-small $N\sim\Delta_\text{charge}\sim t \, e^{-ct/U}$ for some constant $c$. For large $U$, i.e. $U\gg t$, on the other hand
$\Delta_\text{charge}\sim U$. In this limit, it is instructive to expand about $t=0$. Then, the ground state is necessarily insulating: there is one electron
on each site and an energy penalty of $U$ to move any electron since some site would have to be doubly-occupied. For small but non-zero $t/U$,
this picture is dressed by small fluctuations in which an electron makes virtual hops to a neighboring site and then returns. As a result of these fluctuations,
there is an effective interaction between the spins of the electrons on neighboring sites. Since an electron can only hop to a neighboring site if its
spin forms a singlet with the spin of the electron there (due to the Pauli principle), the energy is lowered by such processes if neighboring spins
are antiferromagnetically-correlated. We can derive an effective Hamiltonian at energies much lower than $U$ by taking into account these virtual hopping processes.
It takes the form:
\begin{equation}
H = J \sum_{\langle i, j \rangle} {\bf S}_{i}\cdot {\bf S}_{j} + {\mathcal O}({t^3}/{U^2})
\end{equation}
where $J=4{t^2}/U$. The ${\mathcal O}({t^3}/{U^2})$ terms are due to virtual processes in which multiple hops occcur
.\cite{Thouless65}
If these terms are neglected, this is the antiferromagnetic Heisenberg model.
Its ground state has an antiferromagnetic moment of order $1$. The temperature scale for the onset of antiferromagnetic correlations
is $\sim J$, which is a much lower scale than the charge gap $\sim U$, unlike in the small-$U$ limit, where both scales are comparable.
\footnote{In two dimensions, long-ranged magnetic order is impossible at temperatures $T>0$ and must be stabilized by interactions between
two-dimensional layers. Hence, while the temperature scale at which antiferromegnetic correlations become significant is $\sim J$,
the N\'eel temperature at which long-ranged antiferromagnetic order develops is $\sim J_\perp$, the inter-layer exchange.}
When the subleading ${\mathcal O}({t^3}/{U^2})$ terms are kept, the system remains insulating, but the spin physics may change considerably.
The phase diagram of such spin models is the subject of considerable current research (see, for instance,
Refs. \onlinecite{Misguich99,Melko04}). 

The situation is even less clear when the system is ``doped", i.e. when the density is changed from half-filling.
Then, an effective model can be derived to lowest-order in $t/U$. The model is governed by the $t-J$ Hamiltonian:
\begin{equation}
H = -\sum_{\langle i,j \rangle}\sum_\sigma t_{ij}\left(c^\dag_{i,\sigma}c_{j,\sigma} + \text{h.c.} \right)
+ J \sum_{\langle i, j \rangle} {\bf S}_{i}\cdot {\bf S}_{j} + {\mathcal O}({t^3}/{U^2})
\end{equation}
together with a no-double-occupancy constraint. As a result of this constraint, the $t-J$ model has a smaller per site Hilbert space than the Hubbard model,
which makes it a much more attractive target for exact diagonalization. However,
the $t-J$ model does not capture all of the physics of the Hubbard model (especially in the regime
in which $t/U$ is small but not so small that higher-order in $t/U$ terms can be neglected).

Any deviation from half-filling will cause the conductivity to be non-zero in a completely clean system, in which $\epsilon_i$ is independent of $i$ in Eq. (\ref{eq:hubbard}).
However, in any real material, there will be some disorder due to impurities, and the system will remain insulating for densities very close to half-filling.
As the density deviates from half-filling, antiferromagnetic order is suppressed and eventually disappears. If the Hubbard model captures the physics of the cuprate
superconductors, then superconducting order\footnote{In two dimensions, quasi-long-ranged (i.e. algebraically-decaying) superconducting order can occur at $0<T<T_\text{KT}$ where
$T_\text{KT}$ is the Kosterlitz-Thouless temeprature. At zero temeprature, true long-ranged superconducting order can occur.} with $d_{x^2 - y^2}$ pairing symmetry
must occur for a density of $1-x$ electrons per unit cell, with $x$ in the range
$0.05 \stackrel{<}{\scriptscriptstyle \sim} x \stackrel{<}{\scriptscriptstyle \sim} 0.25$.
Moreover, on the underdoped side of the phase diagram,
$0.05 \stackrel{<}{\scriptscriptstyle \sim}  x \stackrel{<}{\scriptscriptstyle \sim} 0.15$,
other symmetry-breaking order parameters,
such as stripe order, $d$-density wave order, or antiferromagnetic order may also develop  (for recent reviews, see, for instance, Refs. \onlinecite{Sachdev10,Fradkin14}; see, also, Ref. \onlinecite{Corboz14} for a recent calculation finding several
competing orders).
On the other hand, if the Hubbard model is not superconducting in this doping range, then superconductivity in the cuprates must be due to
effects that are missing in the Hubbard model, such as longer-ranged interactions, longer-ranged hopping, interlayer coupling, and phonons.

Thus, as far as the cuprates are concerned, the principle aim of an analysis of the Hubbard model is to determine if it has
a $d_{x^2 - y^2}$ superconducting ground state for $U/t \sim 8$ over the range of dopings
$0.05 \stackrel{<}{\scriptscriptstyle \sim} x \stackrel{<}{\scriptscriptstyle \sim} 0.25$.
If the answer is in the affirmative, then various properties of
this superconducting ground state can be studied. For instance, the gap and its dependence
on the doping, $\Delta(x)$, can be compared to experimental measurements.
If the Hubbard model does not superconduct, then different models must be considered, restoring terms that have been dropped in the passage to the Hubbard model.

\section{Adiabatic preparation of the ground state}
\label{sec:prepare}

We start the adiabatic preparation from the ground state of some Hamiltonian
whose solution is known.  For the sake of concreteness, we consider two examples.
The first, in Sec. \ref{sec:preparemf},  is the quadratic mean-field Hamiltonian whose exact ground state
is the BCS ground state for a $d$-wave superconductor or the analogous Hamiltonian for any
other ordered state such as striped \cite{Salkola96},
antiferromagnetic \cite{Schrieffer88}, $d$-density wave \cite{Nayak00}, etc. states.
The second, discussed in Sec. \ref{sec:plaquettes} is the Hubbard model with hopping matrix elements tuned
so that the system breaks into disconnected $2\times 2$ plaquettes. In either case, the ground state is known and there is a gap to all
excited states. 

An essential point is that these initial states can be efficiently prepared on a quantum computer. Our approach to preparing a Slater determinant, discussed in Sec. \ref{sec:Slater-preparation}, is both deterministic and has better scaling than the algorithm previously proposed in Ref. \onlinecite{Ortiz2001}.

We then adiabatically evolve the Hamiltonian into the Hamiltonian of the Hubbard model with an
additional weak symmetry-breaking term (to give a gap to Goldstone modes, a subtlety that
is discussed in Section \ref{sec:Slater-det}). In Sec. \ref{sec:timeevolve} we review the quantum circuit used to implement time evolution of the Hubbard model and show how it can be done with $\mathcal{O}(N)$ gates and a parallel circuit depth of only $\mathcal{O}(\log N)$ operations per time step.

This step is formally similar to adiabatic quantum optimization \cite{Farhi}, but with a different viewpoint.
If no phase transition occurs along this adiabatic evolution, then we conclude that we guessed correctly about the phase of the system
(for this value of parameters in the final Hubbard model). However, if a phase transition does occur, then the gap will close
and we will know that we guessed incorrectly and can repeat the procedure
with a different initial state in a different phase. Through a process of elimination, we determine the phase of the system
for a given set of parameters and, by repeating this process for different Hubbard model parameters, we map out the phase diagram.
In adiabatic quantum optimzation \cite{Farhi}, it is assumed that a phase transition occurs and it is hoped that the minimum energy
at the transition point scales polynomially in system size rather than exponentially, thereby allowing the evolution to occur in polynomial time.
In our protocol for solving the Hubbard model, however, any gap closing, either polynomial or exponential
-- corresponding, respectively, to a second- or first-order phase transition\cite{sachdev} -- is an invitation to repeat the procedure
with a different initial state and correspondingly different annealing path until no gap closing occurs.
When the initial state is in the same phase as the
ground state of the Hubbard model, the adiabatic evolution can be done in constant time up to subpolynomial factors (see \ref{Sec:anneal} for a more detailed
discussion of errors in adiabatic evolution; while in most of the paper we treat the time required for adiabatic evolution
as scaling with inverse gap squared, in that section we provide a more precise discussion of diabatic errors (i.e., those transitions out of the ground state that are due to a finite time scale for the evolution)
and further show how it is possible to improve the scaling
with gap by using smoother annealing paths.

\subsection{Adiabatic Evolution from Mean-Field States}

\label{sec:preparemf}

\label{sec:Slater-det}

\subsubsection{Mean-Field Hamiltonians}
\label{sec:MF-Hamiltonians}

Let us suppose that we wish to test whether the ground state of the 2D Hubbard model is superconducting for some values of $t, U$ and density.
We can do this by seeing if the Hubbard model is adiabatically connected to some simple superconducting Hamiltonian. It need not be realistic; it merely
needs to be a representative Hamiltonian for a superconductor. We can take the BCS mean-field Hamiltonian:
\begin{eqnarray}
H^{\rm MF}_{DSC}&=&-\sum_{\langle i,j \rangle}\sum_\sigma t_{ij}\left(c^\dag_{i,\sigma}c^{}_{j,\sigma} + c^\dag_{j,\sigma}c^{}_{i,\sigma} \right) \nonumber \\
&& -\sum_{\langle i,j \rangle} \Delta^{{x^2} - y^2}_{ij}\left(c^\dag_{i\uparrow}c^\dagger_{j\downarrow} - c^\dag_{i\downarrow}c^\dagger_{j\uparrow} \right) + \text{h.c.},
\label{eq:mean-field-SC}
\end{eqnarray}
where $\Delta^{{x^2}- y^2}_{ij}=\Delta/2$ for $i=j\pm \hat{\bf x}$ and  $\Delta^{x^2 - y^2}_{ij}=-\Delta/2$ for $i=j\pm \hat{\bf y}$. This is a mean-field $d_{x^2 - y^2}$ superconductor (DSC)
with fixed, non-dynamical superconducting gap of magnitude $\Delta$.

The ground state of this Hamiltonian can be
written in the form of a Slater determinant by making a staggered particle-hole transformation on
the down-spins: $c^\dagger_{i\downarrow} \rightarrow (-1)^{i} c^{}_{i\downarrow}$. Then the
Hamiltonian takes the form:
\begin{multline}
{\tilde H}^{\rm MF}_{DSC} = -\sum_{\langle i,j \rangle}\sum_\sigma t_{ij}\left(c^\dag_{i,\sigma}c^{}_{j,\sigma} + c^\dag_{j,\sigma}c^{}_{i,\sigma} \right) \\
-\sum_{\langle i,j \rangle} (-1)^{j} \Delta^{{x^2} - y^2}_{ij}\left( c^\dagger_{i\uparrow}c^{}_{j\downarrow} - c^\dagger_{j\uparrow} c^{}_{i\downarrow} \right) + \text{h.c.},
\label{eq:mean-field-SC-transformed}
\end{multline}
Under this transformation, the number of up-spin electrons per unit cell is particle-hole transformed:
$n_{i\downarrow}\rightarrow 1- n_{i\downarrow}$ while the number of up-spin electrons is unchanged;
equivalently, the chemical potential for up-spin electrons is flipped in sign, $\mu_\downarrow=-\mu_\uparrow$.
Consequently, the number of down-spin electrons is not equal to the number of up-spin electrons. In addition,
the superconducting pair field $\Delta^{{x^2} - y^2}_{ij}$ has become transformed into a staggered
magnetic field in the $S_x$ direction with a $d_{x^2 - y^2}$ form factor (a $d$-wave spin-density wave, in
the terminology of Ref. \onlinecite{Nayak00}). Thus, the ground state of this quadratic Hamiltonian is
a Slater determinant of the form:
\begin{equation}
|\Psi\rangle = \prod_{\bf k}({u_{\bf k}}c^\dagger_{{\bf k},\uparrow} + {v_{\bf k}}
c^\dagger_{{\bf k}+{\bf Q},\downarrow} )|0\rangle
\end{equation}
Here, the momenta ${\bf k}$ range over the Brillouin zone, $-\frac{\pi}{a} \leq k_{x,y} \leq \frac{\pi}{a}$,
where $a$ is the lattice constant, and the functions ${u_{\bf k}}$, ${v_{\bf k}}$ are
given by:
\begin{eqnarray}
{u^2_{\bf k}} &\equiv& \frac{1}{2}\left(1 - \frac{\epsilon_{k}}{\sqrt{\epsilon^2_{k} + |\Delta_{k}|^2}}\right)\cr
{v^2_{\bf k}} &\equiv& \frac{1}{2}\left(1 + \frac{\epsilon_{k}}{\sqrt{\epsilon^2_{k} + |\Delta_{k}|^2}}\right)
\end{eqnarray}
where $\epsilon_k \equiv -2t(\cos{k_x}a + \cos{k_y}a)$ and
$\Delta_{k} \equiv \Delta(\cos{k_x}a - \cos{k_y}a)$.
In Sec. \ref{sec:Slater-preparation},
we show how to prepare this ground state. For now, we will simply assume that this ground state can be prepared efficiently.
In computing the properties of this state, it is important to keep in mind that physical operators have been transformed according to
$c^\dagger_{i\downarrow} \rightarrow (-1)^{i} c^{}_{i\downarrow}$ and to use the appropriate transformed operators.

We now adiabatically deform this Hamiltonian into a weakly-perturbed Hubbard model. There are two reasons why our final
Hamiltonian is a weakly-perturbed Hubbard model, rather than the Hubbard model itself:
(1) in the absence of long-ranged Coulomb interactions,
(which are neglected in the Hubbard model) a superconducting ground state will have gapless Goldstone excitations,
and (2) a $d_{x^2 - y^2}$ superconductor has gapless fermionic excitations at the
nodes of the superconducting gap. Although these gapless
excitations are important for the phenomenology of superconductors (e.g. the temperature dependence of the superfluid density and
the structure of vortices), they are a nuisance in the present context. Hence, we weakly perturb
the Hubbard model in order to give energy gaps to Goldstone modes and nodal fermions.
However, this can be done weakly so that we do not bias the tendency towards superconductivity
(or lack thereof). If the Hubbard model superconducts, then the scale of the $d_{x^2 - y^2}$ superconducting gap should be
$\sim xJ$, where $\langle n_i \rangle = 1-x$ is the average occupancy of a site and $J=4t^2/U$ is the superexchange interaction. (On the other hand, if the Hubbard model does not superconduct, then the superconducting gap
in the cuprates must be determined by some other energy scale.)
Hence, we apply a fixed $d_{x^2 - y^2}$ superconducting gap that is much smaller than $xJ$ in order to
pin the phase of any superconducting order that may develop. In addition, we apply a small imaginary $d_{xy}$ superconducting
gap to give a small gap to the nodal fermions. Neither gap scales with system size.

More concretely, we adiabatically evolve the Hamiltonian according to:
\begin{multline}
H_{DSC}(s) = (1-s) H^{\rm MF}_{DSC} + s H_{\rm Hub} \\
 -\sum_{\langle i,j \rangle} w \Delta^{{x^2} - y^2}_{ij} \left(c^\dag_{i\uparrow}c^\dagger_{j\downarrow}
 - c^\dag_{i\downarrow}c^\dagger_{j\uparrow} \right) + \text{h.c.}\\
-\sum_{\langle\langle j,k \rangle\rangle} i \Delta^{xy}_{jk}\left(c^\dag_{j\uparrow}c^\dagger_{k\downarrow}
 - c^\dag_{j\downarrow}c^\dagger_{k\uparrow} \right) + \text{h.c.}
\label{eq:mean-field-SC-adiabatic}
\end{multline}
Here, $w$ is a small dimensionless parameter that determines the magnitude of a small pair field that remains throughout the adiabatic evolution.
At $s=0$, it simply increases the superconducting gap by a factor of $1+w$. At $s=1$, it applies a small pair field $\propto w$ to the Hubbard model.
If the Hubbard model does not have a superconducting ground state, then $w\neq 0$ will induce a small superconducting gap $\propto w$.
However, if the Hubbard model does have a superconducting ground state, then $w$ will pin the phase of the superconducting gap,
which would otherwise fluctuate. The magnitude of the gap will be essentially independent of $w$ for small $w$.
If the Hubbard model is superconducting, then the system will remain superconducting all the way from $s=0$ to $s=1$
and the gap will remain open. As long as $s$ is varied slowly compared to the
minimum gap -- which, crucially, is independent of system size -- the system will remain in the ground state.
The final line of Eq. (\ref{eq:mean-field-SC-adiabatic}) is a small imaginary second-neighbor
superconducting gap. Here, $\Delta^{xy}_{jk}=u\Delta/2 $ for $j=k\pm (\hat{\bf x}+\hat{\bf y})$,  $\Delta^{xy}_{jk}=u \Delta/2$ for $j=k\pm (\hat{\bf x}-\hat{\bf y})$,
and $u$ is a small dimensionless number that is independent of system-size. Such a term opens a small gap $\propto u$ at the nodes
(assuming that the Hubbard model does not have a superconducting ground state that spontaneously breaks time-reversal symmetry).
We take $u\Delta \ll xJ$ so that this term opens a gap at the nodes but is otherwise too small to affect the development of superconductivity
away from the nodes.

In summary, Eq. (\ref{eq:mean-field-SC-adiabatic}) is a path through parameter space from a simple superconducting
Hamiltonian with Slater determinant ground state to a Hubbard model that has been perturbed by small terms
that give a gap to order parameter fluctuations and nodal excitations but otherwise do not affect the ground state,
as may be checked by decreasing the magnitude of these perturbations.

A similar strategy can be used to test whether the Hubbard model has any other ordered ground state, such as antiferromagnetic order.
In this case, our starting Hamiltonian is, instead:
\begin{eqnarray}
H^{\rm MF}_{AF}&=&-\sum_{\langle i,j \rangle}\sum_\sigma t_{ij}\left(c^\dag_{i,\sigma}c_{j,\sigma} + c^\dag_{j,\sigma}c_{i,\sigma} \right) \nonumber \\
&& -\sum_{i} (-1)^{i} N  \left(c^\dag_{i\uparrow}c_{i\uparrow} - c^\dag_{i\downarrow}c_{i\downarrow} \right),
\label{eq:mean-field-AF}
\end{eqnarray}
We then evolve along a path analogous to Eq. (\ref{eq:mean-field-SC-adiabatic}), but with the symmetry-breaking terms in the
last two lines replaced by a weak staggered magnetic field. If the Hubbard model has an antiferromagnetic ground state
(for some values of $t, U, x$) then the gap will not close along this path.
However, if the actual ground state of the Hubbard model is superconducting for these values of the couplings, then
the gap will become ${\mathcal O}(1/N^{z/2})$ where $z$ is the dynamical critical exponent of the transition
(i.e. the gap will close, up to finite-size effects)
at the point along the evolution at which superconductivity develops. There will be a non-zero staggered magnetization throughout
since we never fully remove the staggered magnetic field, but there will be a sharp signature at the point at which superconductivity
develops\cite{sachdev}. The presence of a staggered field, which is kept small, may slightly shift the onset of superconductivity but will have no
other effect.

This strategy can allow us, in some of the circumstances in which it has a clear answer,
to address the question: why does the 2D Hubbard model have a superconducting ground
state? In particular, it is possible that the superconducting ground state
of the Hubbard model necessarily has some other type of order coexisting with superconducting order, especially in
the underdoped regime. These secondary orders play a role in some theories of the cuprate superconductors.
If such an order were present in the Hubbard model, evolution from a purely superconducting initial state
would necessarily encounter a phase transition (at the point of onset of this additional order).
Therefore, we can determine not only if the Hubbard model has a superconducting ground state,
but also whether the ground state exhibits a secondary type of long-ranged order over some range of dopings.
We will return to the question ``why does the Hubbard model superconduct'' when we discuss correlation functions.

The virtue of using these mean field Hamiltonians as starting points for adiabatic evolution is that their
physics is fully understood. In particular, their ground states are Slater determinants, which we can efficiently prepare,
as we explain in the next section.

\subsubsection{Preparing Slater Determinants}
\label{sec:Slater-preparation}

In this section, we explain an efficient, simple quantum algorithm to prepare Slater determinants.  This is then used to prepare the Hubbard ground
state by adiabatic evolution from $U=0$ to $U \neq 0$.

The standard algorithm is in Ref.~\onlinecite{Ortiz2001}.  This algorithm suffers from some drawbacks.  Given a projector $\rho$, a Slater determinant state is a state
\be
\Psi_{SD}(\rho)=\Pi_{j=1}^{N_e} b^\dagger_j |0\rangle,
\ee
where $|0\rangle$ is the no particle vacuum state, and each operator $b^\dagger_j$ is a linear combination of $a^\dagger_i$, where the $a^\dagger_i$ are fermion creation operators on a given site $i$.  We write $b^\dagger_j = \sum_{i=1}^n a^\dagger_i P_{ij}$, where $P_{ij}$ are complex scalars and are the entries of $P$, $n$ is the number of orbitals, and $N_e$ is the number of electrons.  Given any  matrix $\rho$, we can find a matrix $P$ such that
\be
\rho=P P^\dagger.
\ee
Then, using this matrix $P$ in the definition of the state gives the state $\Psi_{SD}(\rho)$; note that this definition fixes $\Psi_{SD}(\rho)$ up to a complex phase.
The algorithm in Ref.~\onlinecite{Ortiz2001} proceeds by noting that $U_m=\exp(i \frac{\pi}{2} (b_m+b_m^\dagger))$, acting on the vacuum, produces the state $b_m^\dagger|0\rangle$, up to a phase.  Therefore, if the $b^\dagger_j$ are suitably orthogonalized, successively acting with the unitary $U_m$ produces the desired state.  However, no explicit method for implementing this unitary $U_m$ is given; if the unitary is implemented by a Trotter method, this will be extremely inefficient. We explain a simpler technique.

Let us first assume that $\rho$ is real; the extension to the complex Hermitian case will be straightforward.
The idea is first to prepare the Slater determinant state
\be
\Psi_{SD}(\rho_0) \equiv \Pi_{j=1}^{N_e} a^\dagger_j |0\rangle.
\ee
This corresponds to the case that $\rho$ is an $n$-by-$n$ diagonal matrix, with ones in the first $N_e$ entries and zeroes elsewhere.
This state has one electron on each of the first $N_e$ states.  It is a product state that can be prepared in linear time (indeed, simply initialize each of the first $N_e$ spins to up and the other spins to down).  Then, we act on this state $\Psi_{SD}(\rho)$ with a sequence of {\it Givens rotations}.  A Givens rotation, to borrow the language of matrix 
diagonalization, is a rotation matrix that acts on two rows or columns at a time.  For any given pair, $i,j$, define the operator $R_{i,j}(\theta)$ by
\be
R_{i,j}(\theta) = \exp( \theta c^\dagger_i c_j -h.c.).
\ee
This is a simple two-qubit gate (up to Jordan-Wigner strings) and is, therefore, more straightforwardly implemented
than the unitary $U_m$ in the standard method described above.
Define the $n$-by-$n$ orthogonal matrix $r_{i,j}(\theta)$ by its matrix elements as follows.  Let $(r_{i,j}(\theta))_{k,k}=1$ for $k\neq i,j$ and let $(r_{i,j}(\theta))_{k,l}=0$ if $k \neq l$ and $k \neq i,j$ or if $k \neq l$ and $l \neq i,j$.  Then, let
$(r_{i,j}(\theta))_{i,i}=(r_{i,j}(\theta))_{j,j}=\cos(\theta)$ while $(r_{i,j}(\theta))_{i,j}=-(r_{i,j}(\theta))_{j,i}=\sin(\theta)$.  Then,
\be
R_{i,j}(\theta) \Psi_{SD}(\rho) = \Psi_{SD}(r_{i,j}(\theta) \rho r_{i,j}(\theta)^{-1}),
\ee
where again the equality holds up to the phase ambiguity.

Now, given some desired target $\Psi_{SD}(\rho)$, we claim that one can always find a sequence of at most $n N_e$ Givens rotations that will turn the matrix $\rho$ into the matrix $\rho_0$.  Thus, by inverting this sequence of Givens rotations and applying that sequence to the state $\Psi_{SD}(\rho_0)$, we succeed in constructing the desired state.  The Givens rotations will be ordered such that the Jordan-Wigner strings can be cancelled as in Ref.~\onlinecite{Hastings15}, so the algorithm takes time ${\mathcal O}(N_e n)$.
To show our claim that the sequence of Givens rotations exists, note that the effect of a given rotation which transforms $\rho$ to $r_{i,j}(\theta) \rho r_{i,j}(\theta)^{-1}$ can be achived by transforming $P$ to $r_{i,j}(\theta) P$.  Consider a left singular vector of $P$ with singular value equal to $1$; we can find a sequence of at most $n$ Givens rotations that rotate that vector to be equal to the vector $(1,0,...)$.  As a result, all the other singular vectors now have amplitude only on sites $2,...,n$.  Find another sequence to transform some other left singular vector with singular value equal to $1$ into the vector $(0,1,0,...)$.  Continue until all singular vectors with nonzero singular values have amplitudes only on the first $N_e$ sites.

The extension of this procedure to the complex case is simple.  Rather than having the Givens rotation $R_{i,j}(\theta)$ depends only a single angle $\theta$, we need to perform a rotation $\exp( z c^\dagger_i c_j - h.c.)$ for some complex number $z$.  All the time estimates remain unchanged up to constant factors.

However, for many systems a much shorter sequence of Givens rotations can be found.  Using the same idea as in the fast Fourier transform, if $n$ is a power of $2$, there exists a sequence of $n \log_2(n)$ Givens rotations to transform an $n$-by-$n$ matrix to the momentum basis.  Thus, we can produce ground states of a system of free fermions on a periodic lattice in time ${\mathcal O}(n^2 \log(n))$.  This can be easily extended to handle the case of a periodic system with a unit cell larger than a single site.  Assume there are states $|i,\vec x\rangle$ where $i$ labels an atom in a unit cell and $\vec x$ labels a lattice vector.  We transform to a momentum basis $|i, \vec k\rangle$, and then for each $\vec k$ we initialize some Slater determinant in the unit cell in momentum space.  If the unit cells have size ${\mathcal O}(1)$, the time to initialize each cell is ${\mathcal O}(1)$ so the total time is still ${\mathcal O}(n^2 \log(n))$.

A further possible optimization is that if $N_e>n/2$, we can instead use Givens rotations to bring the holes to the desired states, rather than the electrons, taking a time ${\mathcal O}((n-N_e)n)$ rather than ${\mathcal O}(N_e n)$.  As an interesting application of this, consider creating the ground state of $N_e=3$ electrons in $n=4$ sites, using a hopping Hamiltonian with the sites arranged on a ring.  This can be done by initializing a state with electrons on the first $3$ sites and then apply a total of three different $R_{i,j}(\theta)$ operators.

\subsubsection{Numerical Results}
We applied the Slater determinant preparation procedure to the Hubbard model on $8$ sites arranged in two rows of
$4$ sites.  We started at $U=0$ and $\epsilon_i=0$.  All horizontal couplings were set equal to $1$, while to avoid having an exact ground state degeneracy the vertical couplings were doubled in strength to $2$ (note that this requirement depends on the filling factor; for example, at half-filling we would instead prefer to remain at vertical coupling equal to $1$ to avoid degeneracies).  The ground state of this model is a Slater determinant of free fermion single-particle wavefunctions that we could prepare with a total of $14$ Givens rotations for both spin up and spin down.  This was done by initializing particles in three out of four sites in the top row of sites, and holes in the remaining sites.  Then, $3$ Givens rotations were used to create a uniform superposition of the hole in various states in the top row.  Finally, $4$ Givens rotations were used to create a uniform superposition between top and bottom rows.  This was done for both spins, giving $2 \times 7 = 14$ rotations.

\begin{figure}
\begin{center}
\includegraphics[width=8.5cm]{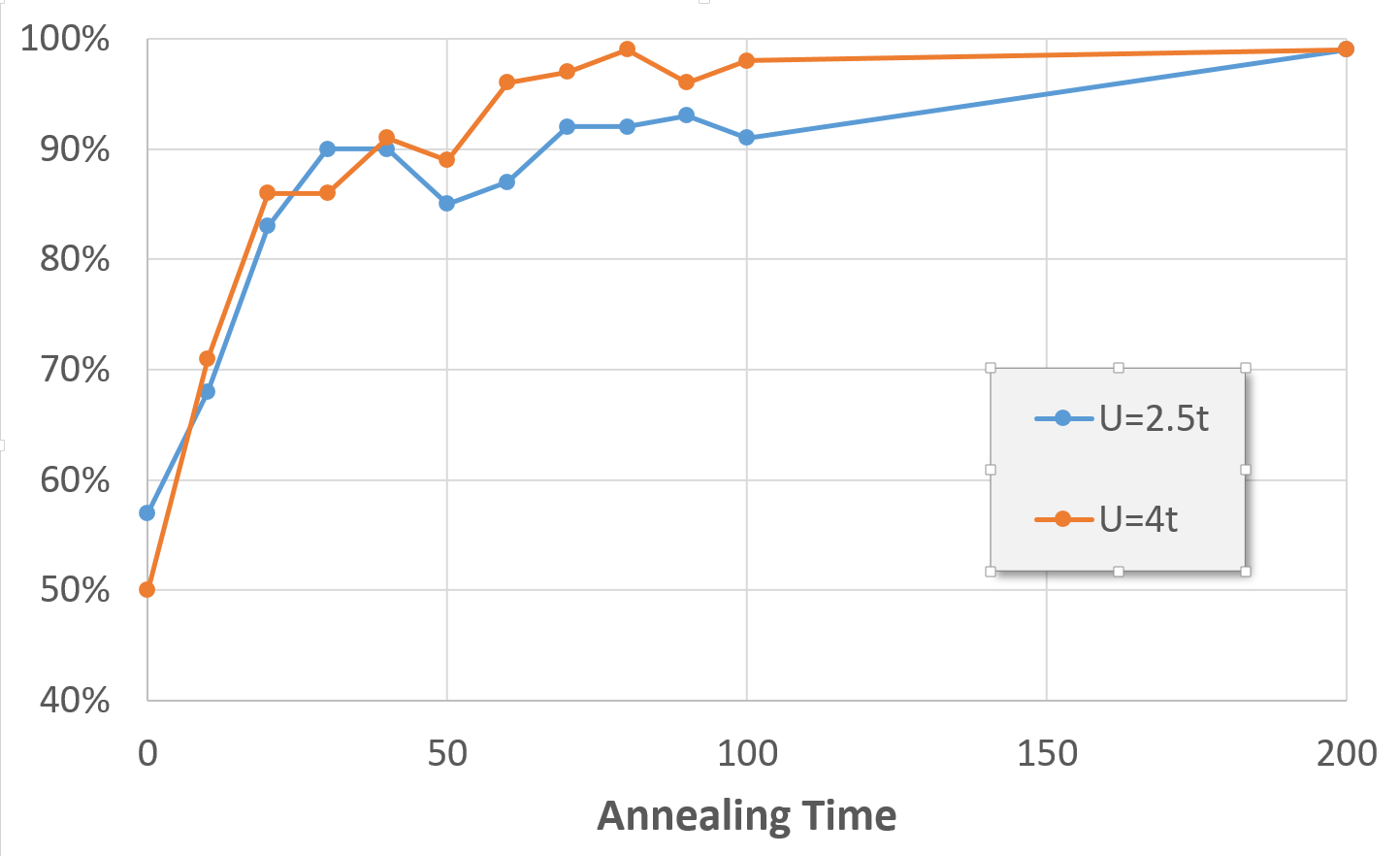}
\end{center}
\caption{Success probability as a function of total annealing time for an annealing path in the Hubbard model
that begins at $U=0$ and vertical hopping equal to $2$ and ends at $U=2.5t$ (blue) or $U=4t$ (red) and all
hoppings equal to $1$. The initial state is a Slater determinant state (of free fermion single-particle states). 
Annealing is linear along the path. The $x$-axis represents the total time for the anneal.  The $y$-axis represents the probability that the phase estimation routine returns an energy that is within $10^{-3}$ of the ground state
energy.  This probability is obtained by sampling runs of the phase estimation circuit, giving rise to some noise in the curves.}
\label{fig:ffanneal}
\end{figure}

We then annealed to $U\neq 0$ while reducing the vertical coupling to unit strength.  The results are shown in Fig.~\ref{fig:ffanneal}.  As seen,
increasing the annealing time increases the success probability, until it converges to $1$.  The annealing times required to achieve substantial (say, $90\%$) overlap with the ground state are not significantly different from the preparation approach discussed next based on ``joining" plaquettes.

Note that for this particular model, we know the ground state energy with very high accuracy from running an exact diagonalization routine so we know whether or not we succeed in projecting onto it after phase estimation.  The system size is small enough that calculating ground state energies is easy to accomplish on a classical computer using Lanczos methods.  Of course, on larger systems requiring a quantum computer to simulate, we will not have
access to the exact energy.  In this case, it will be necessary to do multiple simulations with different mean-field starting states as described at the start of this section; then, if the annealing is sufficiently slow and sufficiently
many samples are taken, the minimum over these simulations will give a good estimate of the ground state energy.  Once we have determined the optimum mean-field starting state and annealing path,
we can then use this path to re-create the ground state to determine correlation functions.

\subsection{Preparing of $d$-wave resonating valence bond states from plaquettes}
\label{sec:plaquettes}

\subsubsection{ $d$-wave resonating valence bond states}
An alternative approach to adiabatic state preparation has previously been proposed in the context of ultracold quantum gases in optical lattices. In this approach, $d$-wave resonating valence bond (RVB) states are prepared.\cite{Trebst2005}
RVB states are spin liquid states of a Mott insulator, in which the spins on nearby sites
are paired into short-range singlets.
RVB states were first introduced as proposed wave functions for spin liquid
ground states of Mott insulators\cite{Anderson:1973eo}, and then conjectured to describe the insulating parent
compounds of the cuprate superconductors.\cite{Anderson:1987ii} In more modern language, they
are described as $\mathbb{Z}_2$ topologically-ordered spin liquid states.\cite{Wen91c}
The idea was that, upon doping, the
singlet pairs would begin to move and form a condensate of Cooper pairs. Although the insulating parent compounds
of the cuprates are antiferromagnetically-ordered, it is nevertheless possible that the cuprate superconductors are close to
an RVB spin liquid ground state and are best understood as doped spin liquids.
The RVB scenario has been confirmed for $t$-$J$ and Hubbard models of coupled plaquettes
\cite{fye,Altman:2002iu} and ladders
(consisting of two coupled chains) \cite{Tsunetsugu:94,Troyer:96} and remains a
promising candidates for the ground state of the Hubbard model and high-temperature superconductors.\cite{RVB:04}

Instead of starting from mean-field Hamiltonians, we build up the ground state wave function from small spatial motifs, in this case four-site plaquettes, which are the smallest lattice units on which electrons pair in the Hubbard model.\cite{fye} Following the same approach as originally proposed for ultracold quantum gases, we prepare four-site plaquettes in their ground state, each filled with either two or four electrons. \cite{Trebst2005} These plaquettes then get coupled, either straight to a two-dimensional square lattice, or first to quasi-one dimensional ladders, which are subsequently coupled to form a square lattice.

\subsubsection{Preparing the ground states of four-site plaquettes}

\begin{figure}
\begin{center}
\includegraphics[width=3cm]{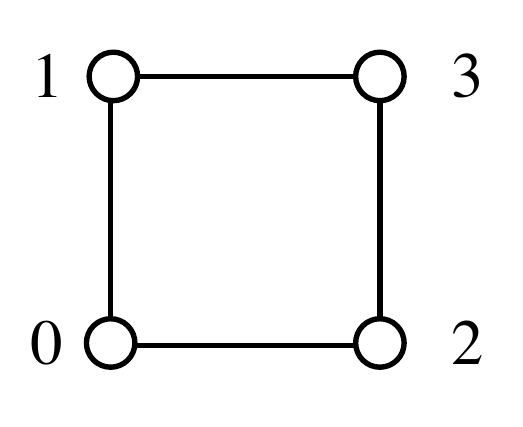}
\end{center}
\caption{Labeling of the sites in a plaquettes}
\label{fig:plaquette}
\end{figure}

The first step is preparing the ground state of the Hubbard model on 4-sites plaquettes filled with either two or four electrons. We start from very simple product states and adiabatically evolve them into the ground states of the plaquettes.

To prepare the ground states of plaquettes with four electrons we start from a simple product state $c^\dag_{0,\uparrow}c^\dag_{0,\downarrow}c^\dag_{1,\uparrow}c^\dag_{1,\downarrow}|0\rangle$, using the labeling shown in Fig. \ref{fig:plaquette}. Our initial Hamiltonian, of which this state is the ground state has no hopping ($t_{ij}=0$), but already includes the Hubbard repulsion on all sites. To make the state with doubly occupied state the ground state we add large on-site potentials $\epsilon_2 = \epsilon_3 = 2U+3t$ on the empty sites. To prepare the ground state of the plaquette we then 
\begin{enumerate}
\item start with $t_{ij}=0$, $U_i=U$ and a large potential on the empty sites  (here $\epsilon_2 = \epsilon_3 = \epsilon$) so that the initial state is the ground state,
\item ramp up the hoppings $t_{ij}$ from 0 to $t$ during a time $T_1$,
\item ramp  the potentials $\epsilon_{i}$ down to 0 during a  time $T_2$.
\end{enumerate}
Time scales $T_1=T_2\approx10 t^{-1}$ and $\epsilon=U+4t$ are sufficient to achieve high fidelity.
As we discuss below, it can be better to prepare the state quickly and then project into the ground state through a quantum phase estimation than to aim for an extremely high fidelity in the adiabatic state preparation.
 To prepare a plaquette with two electrons we start from the  product state $c^\dag_{0,\uparrow}c^\dag_{0,\downarrow}|0\rangle$ and use the same schedule, except that we choose the non-zero potentials on three sites of the plaquette ($\epsilon_1= \epsilon_2 = \epsilon_3 = \epsilon$).

\subsubsection{Coupling of plaquettes}

\begin{figure}
\begin{center}
\includegraphics[width=6cm]{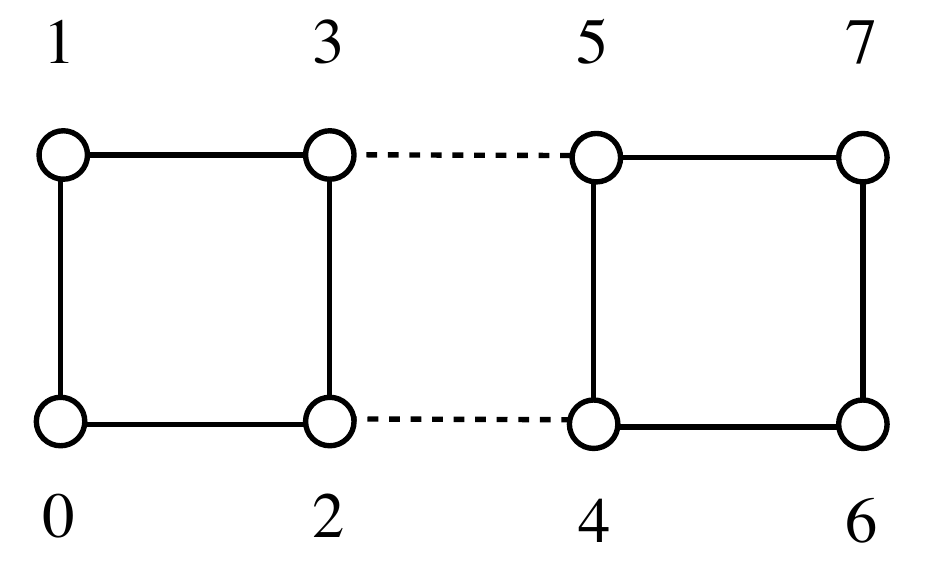}
\end{center}
\caption{Coupling of two plaquettes}
\label{fig:plaquettes}
\end{figure}

After preparing an array of decoupled plaquettes, some of them with four electrons and some with two electrons depending on the desired doping, we adiabatically turn on the coupling between the plaquettes. As a test case we use that of Ref. \onlinecite{Trebst2005} and couple two plaquettes (see Fig. \ref{fig:plaquettes}), one of them prepared with two electrons and one with four electrons.

Naively coupling plaquettes by adiabatically turning on the intra-plaquette couplings $t_{24}$ and  $t_{35}$ will not give the desired result, since the Hamiltonian is reflection symmetric but an initial state with two electrons on one plaquette and four on the other breaks this symmetry. No matter how slowly the anneal is done, the probability of preparing the ground state will not converge to $1$.  We thus need to explicitly break the reflection symmetry by either first ramping up a small chemical potential on a subset of the sites or -- more easily -- not completely ramping down the non-zero $\epsilon_i$ values when preparing plaquette ground states. Consistent with Ref. \onlinecite{Trebst2005} we find that a time $T_3\approx50t^{-1}$ is sufficient to prepare the ground state with high fidelity. 

\begin{figure}
\begin{center}
\includegraphics[width=8.5cm]{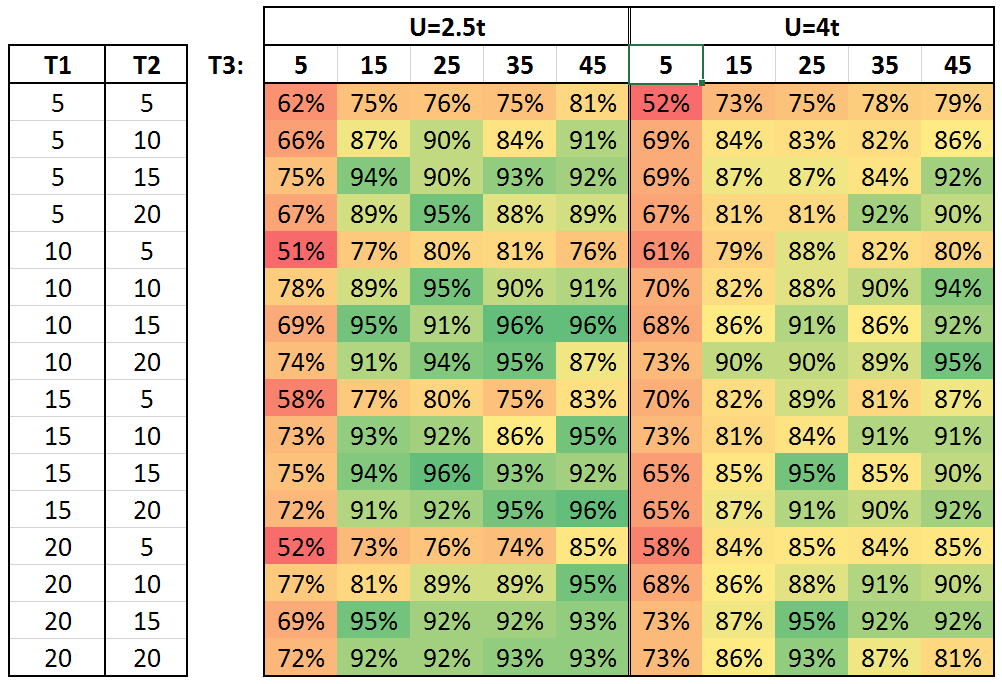}
\end{center}
\caption{Probability of preparing ground state as a function of times $T_1$, $T_2$ ,$T_3$ in units of $t^{-1}$, following the annealing schedule described above.  During time $T_3$, a uniform potential $\epsilon$ was ramped from $t$ to $0$ on the plaquette with two electrons to break the symmetry between plaquettes.}
\label{fig:annealjoin}
\end{figure}

\begin{figure}
\begin{center}
\includegraphics[width=8.5cm]{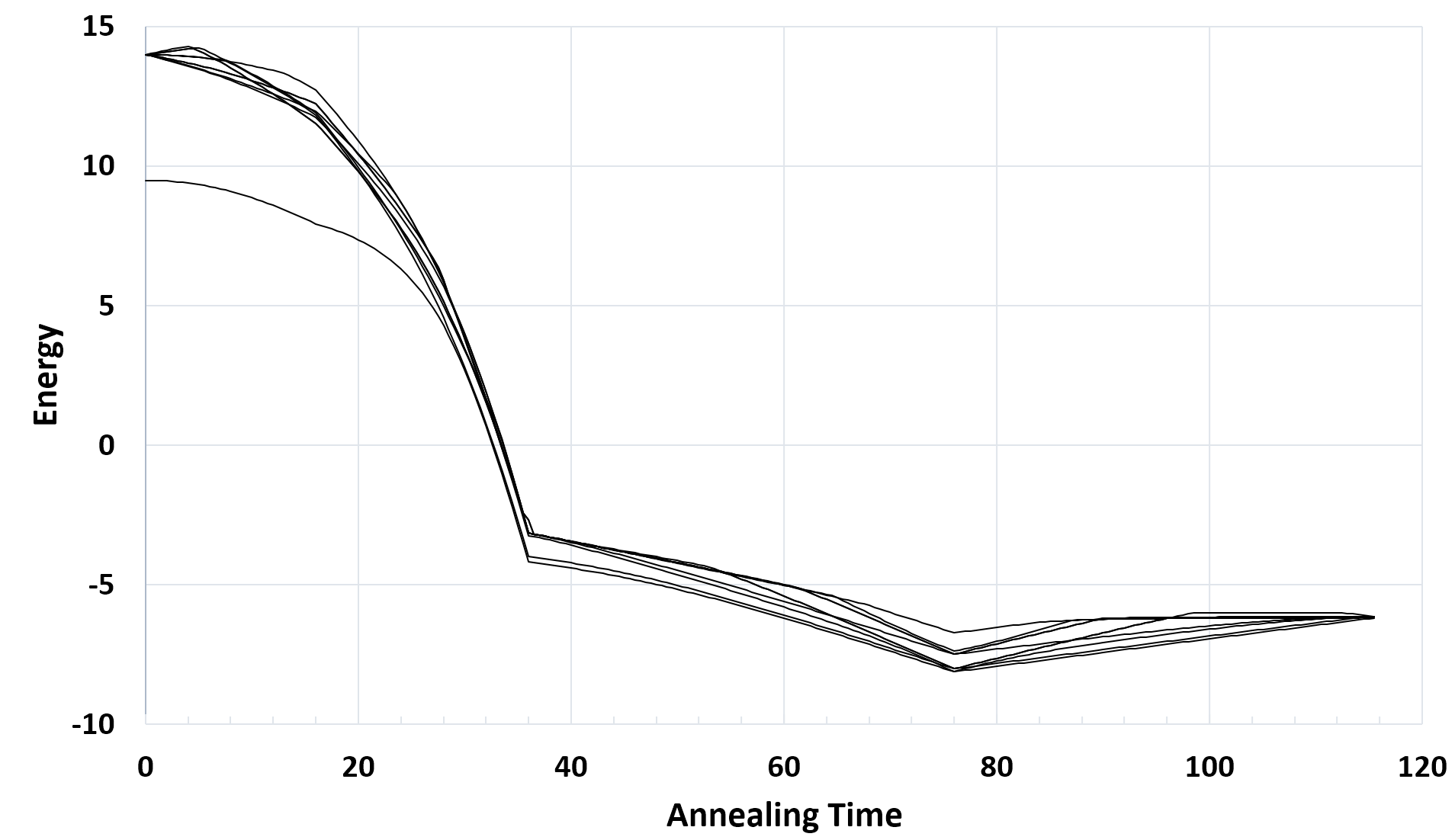}
\end{center}
\caption{Energy levels during annealing.  $y$-axis is energy.  $x$ axis is time.  From time $0$ to time $16$ (in units of $t^{-1}$) the hopping is ramping up, time $16$ to time $36$ the non-uniform $\epsilon$ in each plaquette is ramping down (leaving a uniform $\epsilon=1$ on the plaquette with two electrons), time $36$ to time $76$ is joining plaquettes.  All levels remain non-degenerate except for a degeneracy that appears at the end of the anneal.  After time $76$ plaquettes are separated, as described below.}
\label{fig:gapsjoin}
\end{figure}

See Fig.~\ref{fig:annealjoin} for success probabilities depending on times $T_1$ to ramp up hopping in a plaquette, $T_2$ to ramp down $\epsilon$ in a plaquette, and $T_3$ to join the plaquettes.
The specific couplings in these figures are an initial potential $8t$ on the  empty sites in the plaquette with four electrons, chemical potential $8t$ on the empty sites in the plaquette with two electrons and chemical potential  $t$ on the occupied site in the plaquette with two electrons.  See Fig.~\ref{fig:gapsjoin}
for a plot of the spectral gaps for an example annealing schedule.

Going from two plaquettes to the full square lattice is straightforward: we bias some plaquettes with small values of $\epsilon_i$ to break any reflection and rotation symmetries, and then switch on inter-plaquette hoppings, and switch off the bias potentials. The time scale here is a-priori unknown. According to Ref. \onlinecite{Trebst2005} the ground state on ladders can be prepared within a relatively short time $T\approx200t^{-1}-500t^{-1}$. In two dimensions we expect that similar time scales may be sufficient unless we encounter a quantum phase transition to a different phase.

\subsubsection{Decoupling plaquettes}

As proposed in the context of analog quantum simulation, the pairing of holes on plaquettes can be tested by adiabatically decoupling plaquettes.\cite{Trebst2005} We start by preparing one plaquette with four electrons and a second plaquette with  two electrons.  We then couple them as described above to end up with a system with two holes relative to the half filled Mott insulator. After adiabatically decoupling the two plaquettes again we measure the number of electrons on each plaquette. If holes bind in the ground state, they will end up  on the same plaquette, and we measure an even electron number per plaquette. If they don't bind each hole will prefer to be on a different plaquette to maximize its kinetic energy and we will end up with three electrons per plaquette.

\begin{figure}
\begin{center}
\includegraphics[width=8.5cm]{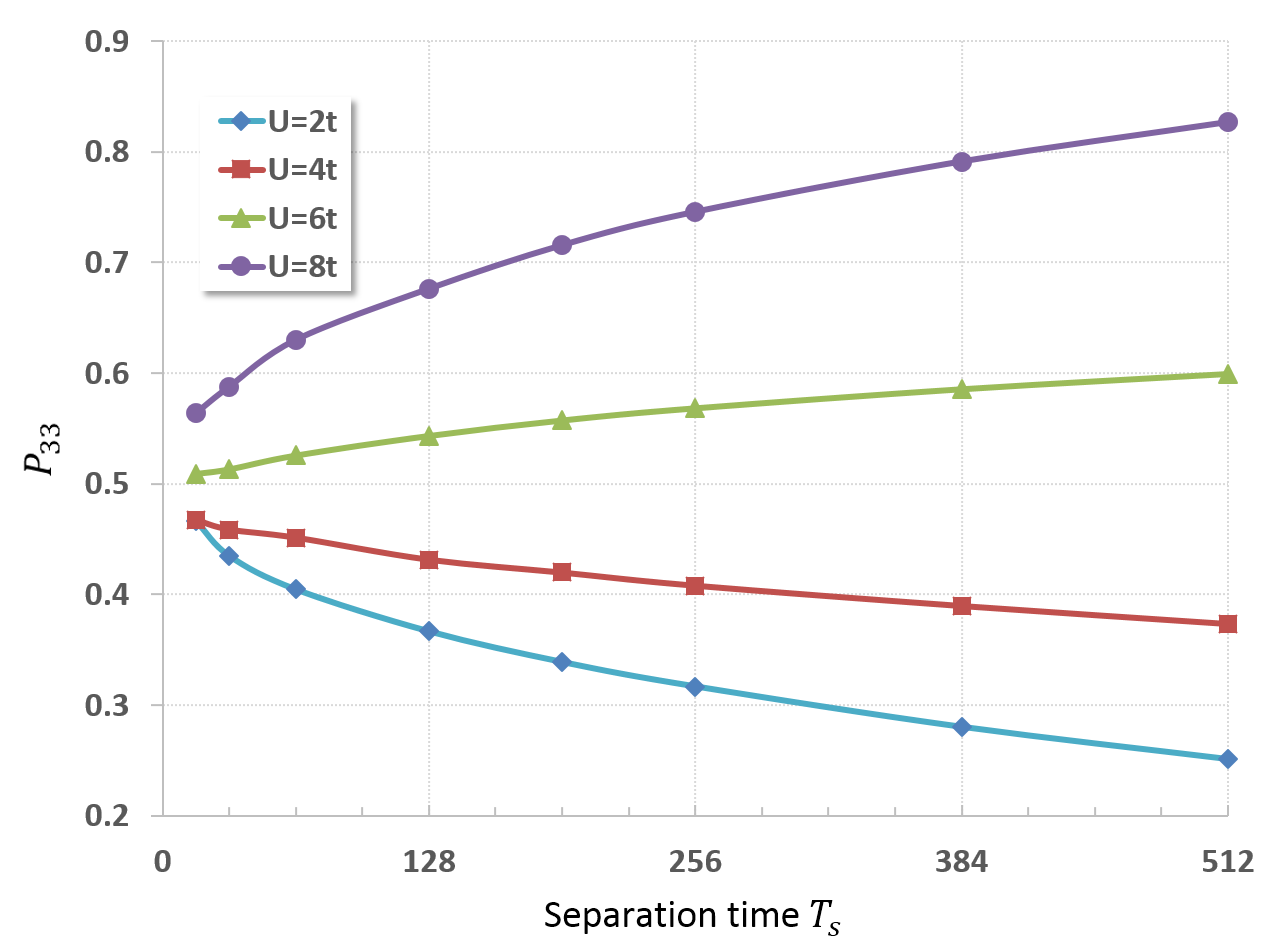}
\end{center}
\caption{Shown is the probability $P_{33}$ of finding three electrons on each of the plaquette after decoupling two plaquettes with a total of six electrons as a function of the time $T_s$ (in units of $t^{-1}$) over which the plaquettes were nearly adiabatically separated. A decrease of   $P_{33}$ indicates pairing of holes in the ground state of a plaquette.}
\label{fig:sep}
\end{figure}

As shown in Fig. \ref{fig:sep}, the probability of finding three electrons on a plaquette goes to zero for $U=2t$ and $U=4t$ when increasing the time $T_s$ over which we ramp down the inter-plaquette hopping. This indicates pairing and is consistent with the known critical value of $U\approx4.5t$ for pairing on four-site plaquettes. On the other hand for $U=6t$ and $U=8t$ we see an increase in the probability of finding three electrons per plaquette, indicating the absence of pairing.

\subsection{Trotter Errors and Accelerating adiabatic preparation by using larger time steps}
In practice, it suffices for our annealing preparation to obtain a state with sufficiently large overlap onto the ground state that phase estimation will then project onto the ground state with a reasonably large probability.  We can thus tolerate some errors in our annealing path.  For this reason, we annealed using a larger time step, before doing the phase estimation with a smaller time step.

In fact, this approach of annealing at a larger time step can be turned into an approach which uses a larger time step but still produces a state with very large overlap with the ground state.  Note that the larger Trotter-Suzuki step means that we in fact evolve under a modified Hamiltonian, proportional to the logarithm of the unitary describing a single time step.  As the time step decreases, this Hamiltonian becomes closer to the desired Hamiltonian.  It is then possible to evolve with a larger time step, following the adiabatic path in parameter space, and then at the end of the path one can ``anneal in the  time step", gradually reducing the time step to zero.

In our simulation of the anneal using joining two plaquettes, the system was rather insensitive to Trotter errors.  
We used a relatively large time step, equal to $0.25t^{-1}$ for the annealing, and used a much smaller time step for the phase estimation that made the errors negligible there. 

A time step of $0.05t^{-1}$ for phase estimation yielded relative errors below $10^{-3}$ and $0.003t^{-1}$ yielded relative errors below $10^{-5}$.
The larger time step used for the adiabatic preparation  prevents us from obtaining $100\%$ overlap with the ground state even in the limits of very long annealing times; however, as seen we still obtain very high overlaps in the range of $80\%-96\%$.

\section{Implementing unitary time evolution of the Hubbard model}

\label{sec:timeevolve}

\subsection{Circuits for the individual terms}

For the Hubbard model, we need to implement unitary evolution under two types of terms, hopping terms $c^\dagger_{p, \sigma} c_{q,\sigma}$, and repulsion terms $c^\dagger_{p,\sigma} c_{p,\sigma} c^\dagger_{q,\sigma} c_{q,\sigma}$.  These terms are a subset of those needed for earlier electronic structure calculations\cite{whitfield2011}. In addition we will need to implement unitary evolution under a pairing term $c_{p,\sigma}^\dag c_{q,\sigma'}^\dag+h.c.$. 

\begin{table}
\begin{tabular}{|c|c|c|}
\hline 
Name & Symbol & Matrix Representation \\
\hline
\hline
Hadamard & 
\begin{tikzpicture}[scale=0.50,every node/.style={scale=0.50}]
\matrix[matStyle] {
  \point[0-0] & \gate[0-1]{H}) & \point[0-2]  \\
};
\begin{pgfonlayer}{background}
\qw{0-0}{0-2} 
\end{pgfonlayer}
\end{tikzpicture}
& 
$\left[ \begin{array}{cc} 1/\sqrt{2} &1/\sqrt{2} \\ 1/\sqrt{2} & -1/\sqrt{2} \end{array} \right]$ \\
\hline 

$Y$ basis change & 
\begin{tikzpicture}[scale=0.50,every node/.style={scale=0.50}]
\matrix[matStyle] {
  \point[0-0] & \gate[0-1]{Y}) & \point[0-2]  \\
};
\begin{pgfonlayer}{background}
\qw{0-0}{0-2} 
\end{pgfonlayer}
\end{tikzpicture}
& 
$\left[ \begin{array}{cc} 1/\sqrt{2} &i/\sqrt{2} \\ i/\sqrt{2} & 1/\sqrt{2} \end{array} \right]$ \\
\hline 

Phase gate & 
\begin{tikzpicture}[scale=0.50,every node/.style={scale=0.50}]
\matrix[matStyle] {
  \point[0-0] & \gate[0-1]{T(\theta)}) & \point[0-2]  \\
};
\begin{pgfonlayer}{background}
\qw{0-0}{0-2} 
\end{pgfonlayer}
\end{tikzpicture}
& 
$\left[ \begin{array}{cc} 1 &0 \\ 0 & e^{-i\theta} \end{array} \right]$ \\ 
\hline

$Z$ rotation  & 
\begin{tikzpicture}[scale=0.50,every node/.style={scale=0.50}]
\matrix[matStyle] {
  \point[0-0] & \gate[0-1]{R_z(\theta)}) & \point[0-2]  \\
};
\begin{pgfonlayer}{background}
\qw{0-0}{0-2} 
\end{pgfonlayer}
\end{tikzpicture}
& 
$\left[ \begin{array}{cc} e^{i\theta/2} &0 \\ 0 & e^{-i\theta/2} \end{array} \right]$ \\ 
\hline

controlled not (CNOT)  & 
\begin{tikzpicture}[scale=0.7,every node/.style={scale=0.7}]
\matrix[matStyle] {
  \point[0-0] & \ctrl[0-1] & \point[0-2]  \\
  & & \\
  \point[1-0] & \targ[1-1] & \point[1-2]  \\
};
\begin{pgfonlayer}{background}
\qw{0-0}{0-2} 
\qw{1-0}{1-2} 
\qwx{0-1}{1-1}
\end{pgfonlayer}
\end{tikzpicture}
& 
$\left[ \begin{array}{cccc} 1 & 0 & 0 & 0 \\ 0 & 1 & 0 & 0 \\ 0 & 0 & 0 & 1 \\ 0 & 0 & 1 & 0 \end{array} \right]$ \\ 
\hline

\end{tabular}
\caption{An overview of the quantum gates used in this paper. Note that $Y$ gates do not represent Pauli $Y$; rather, they represent a Clifford operator that interchange $Y$ and $Z$ spin orientations. The phase gate $T(\theta)$ can be replaces by an $R_z(\theta)$ rotation, up to a global phase that can be kept track of classically.}
\label{tab:gates}
\end{table}

To establish notation we summarize In Table \ref{tab:gates} the gates used in the quantum circuits shown in this paper.

\subsubsection{The repulsion and chemical potential terms}

\begin{figure}
\begin{tikzpicture}[scale=0.70,every node/.style={scale=0.70}]
\matrix[matStyle] {
  \point[0-0] & & \gate[0-2]{R_{z}(\theta/2)}  & \ctrl[0-3] &  & \ctrl[0-5] & \point[0-6]  \\
  \point[1-0] &  &  &  &  &  & \point[1-6]  \\
  \point[2-0] &  &  &  &  &  & \point[2-6]  \\
  \point[3-0] &  &  &  &  &  & \point[3-6]  \\
  \point[4-0] &  &  &  &  &  & \point[4-6]  \\
  \point[5-0] &  & \gate[5-2]{R_{z}(\theta/2)} & \targ[5-3] & \gate[5-4]{R_{z}(-\theta/2)} & \targ[5-5] & \point[5-6]  \\
   &  &  &  &  &  &   \\
};
\begin{pgfonlayer}{background}
\qw{0-0}{0-6} \qw{1-0}{1-6} \qw{2-0}{2-6} \qw{3-0}{3-6} \qw{4-0}{4-6} \qw{5-0}{5-6} 

\qwx{0-3}{5-3} \qwx{0-5}{5-5} 
\end{pgfonlayer}
\end{tikzpicture}
\caption{\label{fig:Hpqqp} Quantum circuit to implement time evolution for a time step $\theta$ under the term $H_{pqqp} = n_{p,\sigma} n_{q,\sigma'} = c^\dagger_{p,\sigma}c^\dagger_{q,\sigma'} c_{q,\sigma'} c_{p,\sigma}$.
  Top and bottom lines represent qubits for $p,\sigma$ and $q,\sigma'$. }
\end{figure}
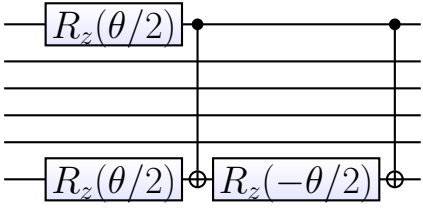

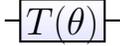
\begin{figure}
\begin{tikzpicture}[scale=0.70,every node/.style={scale=0.70}]
\matrix[matStyle] {
  \point[0-0] & \gate[0-1]{T(\theta)}) & \point[0-2]  \\
   &  &   \\
};
\begin{pgfonlayer}{background}
\qw{0-0}{0-2} 
\end{pgfonlayer}
\end{tikzpicture}
\caption{\label{fig:Hpp} Quantum circuit to implement time evolution for a time step $\theta$ under the  chemical potential term $n_{p,\sigma} =c^\dagger_{p,\sigma} c_{p,\sigma}$. }
\end{figure}

The repulsion term is an example of what is called an $H_{pqqp}$ term and evolution under this term is given in Fig. \ref{fig:Hpqqp}.
A chemical potential term, $c^\dagger_{p,\sigma} c_{p,\sigma}$, which we call an $H_{pp}$ term, is not needed for the final Hamiltonian but is useful for annealing.  The circuit for this term
is shown in Fig. \ref{fig:Hpp}.
Since all the repulsion terms in the Hubbard model and all the chemical potential terms commute with each other, they can be applied in parallel, thus needing only $\mathcal{O}(N)$ gates and $\mathcal{O}(1)$ parallel circuit depth.

\subsubsection{The hopping and pairing terms}

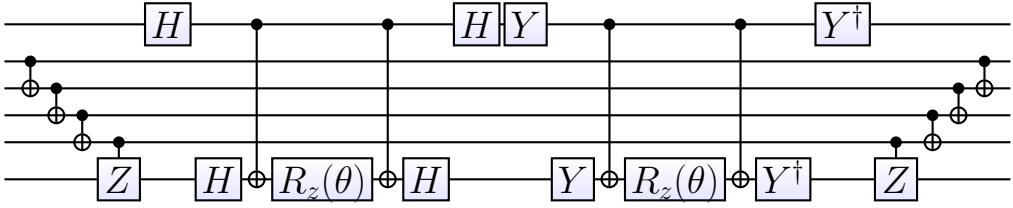
\begin{figure*}
\begin{tikzpicture}[scale=0.70,every node/.style={scale=0.70}]
\matrix[matStyle] {
  \point[0-0] &  &  &  &  & \gate[0-5]{H} &  & \ctrl[0-7] &  & \ctrl[0-9] &  & \gate[0-11]{H} & \gate[0-12]{Y} &  & \ctrl[0-14] &  & \ctrl[0-16] &  & \gate[0-18]{Y^\dagger} &  &  &  &  & \point[0-23]  \\
  \point[1-0] & \ctrl[1-1] &  &  &  &  &  &  &  &  &  &  &  &  &  &  &  &  &  &  &  &  & \ctrl[1-22] & \point[1-23]  \\
  \point[2-0] & \targ[2-1] & \ctrl[2-2] &  &  &  &  &  &  &  &  &  &  &  &  &  &  &  &  &  &  & \ctrl[2-21] & \targ[2-22] & \point[2-23]  \\
  \point[3-0] &  & \targ[3-2] & \ctrl[3-3] &  &  &  &  &  &  &  &  &  &  &  &  &  &  &  &  & \ctrl[3-20] & \targ[3-21] &  & \point[3-23]  \\
  \point[4-0] &  &  & \targ[4-3] & \ctrl[4-4] &  &  &  &  &  &  &  &  &  &  &  &  &  &  & \ctrl[4-19] & \targ[4-20] &  &  & \point[4-23]  \\
  \point[5-0] &  &  &  & \gate[5-4]{Z} &  & \gate[5-6]{H} & \targ[5-7] & \gate[5-8]{R_{z}(\theta)} & \targ[5-9] & \gate[5-10]{H} &  &  & \gate[5-13]{Y} & \targ[5-14] & \gate[5-15]{R_{z}(\theta)} & \targ[5-16] & \gate[5-17]{Y^\dagger} &  & \gate[5-19]{Z} &  &  &  & \point[5-23]  \\
   &  &  &  &  &  &  &  &  &  &  &  &  &  &  &  &  &  &  &  &  &  &  &   \\
};
\begin{pgfonlayer}{background}
\qw{0-0}{0-23} \qw{1-0}{1-23} \qw{2-0}{2-23} \qw{3-0}{3-23} \qw{4-0}{4-23} \qw{5-0}{5-23} 

\qwx{0-7}{5-7} \qwx{0-9}{5-9} \qwx{0-14}{5-14} \qwx{0-16}{5-16} \qwx{1-1}{2-1} \qwx{1-22}{2-22} 
\qwx{2-2}{3-2} \qwx{2-21}{3-21} \qwx{3-3}{4-3} \qwx{3-20}{4-20} \qwx{4-4}{5-4} \qwx{4-19}{5-19} 

\end{pgfonlayer}
\end{tikzpicture}

\caption{\label{fig:Hpq} Quantum circuit to implement time evolution for a time step $\theta$ under the  hopping term $c_{p,\sigma}^\dagger c_{q,\sigma}+c_{q,\sigma}^\dagger c_{p,\sigma}$. Top and bottom lines represent qubits for $p,\sigma$ and $q,\sigma$. }
\end{figure*}

The unitary evolution under the hopping term, that we call an $H_{pq}$ term, is given by the circuit in Fig. \ref{fig:Hpq}, and has also been considered
previously.
It also will be useful to be able to implement evolution under a super-conducting pairing term, 
$\Delta c_{p,\sigma}^\dag c_{q,\sigma'}^\dag+h.c.$.  This term is similar to the hopping term, since both are quadratic in the fermionic operators.
The hopping and pairing term circuits will be very similar to each other, with the changes involving changing some of the basis 
change gates.
The pairing term will be needed for the adiabatic evolution from the BCS mean-field Hamiltonian suggested in subsection 
\ref{sec:MF-Hamiltonians}.  Also, it will be needed to measure superconducting correlations present in the Hubbard state: in this case the measurement will be of the correlation function between two different superconducting pairing terms on two pairs of sites separated from each other.  Measurement will be discussed more later, but our general procedure for measurement will require knowledge of the unitary which implements evolution under the term.

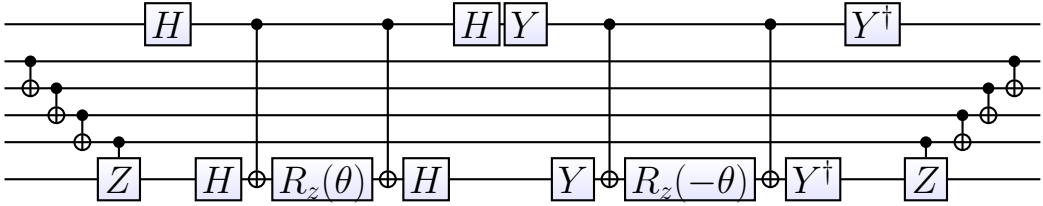
\begin{figure*}
\begin{tikzpicture}[scale=0.70,every node/.style={scale=0.70}]
\matrix[matStyle] {
  \point[0-0] &  &  &  &  & \gate[0-5]{H} &  & \ctrl[0-7] &  & \ctrl[0-9] &  & \gate[0-11]{H} & \gate[0-12]{Y} &  & \ctrl[0-14] &  & \ctrl[0-16] &  & \gate[0-18]{Y^\dagger} &  &  &  &  & \point[0-23]  \\
  \point[1-0] & \ctrl[1-1] &  &  &  &  &  &  &  &  &  &  &  &  &  &  &  &  &  &  &  &  & \ctrl[1-22] & \point[1-23]  \\
  \point[2-0] & \targ[2-1] & \ctrl[2-2] &  &  &  &  &  &  &  &  &  &  &  &  &  &  &  &  &  &  & \ctrl[2-21] & \targ[2-22] & \point[2-23]  \\
  \point[3-0] &  & \targ[3-2] & \ctrl[3-3] &  &  &  &  &  &  &  &  &  &  &  &  &  &  &  &  & \ctrl[3-20] & \targ[3-21] &  & \point[3-23]  \\
  \point[4-0] &  &  & \targ[4-3] & \ctrl[4-4] &  &  &  &  &  &  &  &  &  &  &  &  &  &  & \ctrl[4-19] & \targ[4-20] &  &  & \point[4-23]  \\
  \point[5-0] &  &  &  & \gate[5-4]{Z} &  & \gate[5-6]{H} & \targ[5-7] & \gate[5-8]{R_{z}(\theta)} & \targ[5-9] & \gate[5-10]{H} &  &  & \gate[5-13]{Y} & \targ[5-14] & \gate[5-15]{R_{z}(-\theta)} & \targ[5-16] & \gate[5-17]{Y^\dagger} &  & \gate[5-19]{Z} &  &  &  & \point[5-23]  \\
   &  &  &  &  &  &  &  &  &  &  &  &  &  &  &  &  &  &  &  &  &  &  &   \\
};
\begin{pgfonlayer}{background}
\qw{0-0}{0-23} \qw{1-0}{1-23} \qw{2-0}{2-23} \qw{3-0}{3-23} \qw{4-0}{4-23} \qw{5-0}{5-23} 

\qwx{0-7}{5-7} \qwx{0-9}{5-9} \qwx{0-14}{5-14} \qwx{0-16}{5-16} \qwx{1-1}{2-1} \qwx{1-22}{2-22} 
\qwx{2-2}{3-2} \qwx{2-21}{3-21} \qwx{3-3}{4-3} \qwx{3-20}{4-20} \qwx{4-4}{5-4} \qwx{4-19}{5-19} 

\end{pgfonlayer}
\end{tikzpicture}
\caption{\label{fig:HpqPair} Quantum circuit to implement time evolution for a time step $\theta$ under the  pairing term $c_{p,\sigma}^\dagger c_{q,\sigma'}^\dagger+c_{q,\sigma'} c_{p,\sigma}$. Top and bottom lines represent qubits for $p,\sigma$ and $q,\sigma'$.}
\end{figure*}

For the case $\Delta$ real, we wish to implement unitary evolution by $\exp[-i \theta (c_{p,\sigma}^\dag c_{q,\sigma'}^\dag+h.c.)]$.  Let us write $X_{p,\sigma}$ to denote the Pauli $X$ operator on the qubit corresponding to spin-orbital $p,\sigma$, and similarly $Y_{p,\sigma}$ and $Z_{p,\sigma}$ denote Pauli $Y$ and $Z$ operators on that spin-orbital.
Using the identity that $c^\dagger=(1/2)(X+iY)$ and $c=(1/2)(X-iY)$, up to phase factor associated with the Jordan-Wigner strings,  
we wish to implement
\begin{equation}
\exp[-i \frac{\theta}{2}(X_{p,\sigma} X_{q,\sigma}-Y_{p,\sigma} Y_{q,\sigma'}) Z_{JW}],
\end{equation}
where $Z_{JW}$ denotes the product of the Pauli $Z$ operators on the spin-orbitals intervening between $p,\sigma$ and $q,\sigma'$.  In general, we wish to implement this unitary operation controlled by some ancilla.
In fact, this circuit is the {\it same} as the circuit used to implement the term $c^\dagger_{p,\sigma} c_{q,\sigma'}+h.c.$, up to a sign change on the second spin rotation, as shown in Fig. \ref{fig:HpqPair} .

For $\Delta$ imaginary, we wish to implement 
\begin{equation}
\exp[-i \frac{\theta}{2}(X_{p,\sigma} Y_{q,\sigma}+Y_{p,\sigma} X_{q,\sigma'}) Z_{JW}].
\end{equation}
This circuit is similar to the other pairing and hopping circuits, except that in the first half of the circuit, we apply a Hadamard to the first qubit and
a $Y$ basis change to the second, and in the  second half we apply the $Y$ basis change to the first qubit and the Hadamard to the second.

\subsection{Optimizing the cost of the Jordan Wigner transformation}
\label{optcost}
\begin{figure}
\begin{center}
\includegraphics[width=6cm]{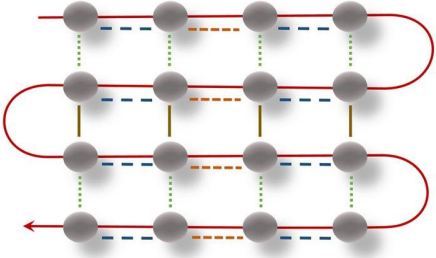}
\end{center}
\caption{Snake pattern for ordering of sites.  Arrow shows order of sites.  Dashed and solid lines represent hopping terms.  The hopping terms are grouped into four different sets, two sets of vertical and two sets of horizontal hoppings as shown by different colorations and dash patterns.}
\label{fig:snake}
\end{figure}

In order to implement the unitary time evolution in the Hubbard model, we use a Jordan-Wigner transformation to turn the model into a spin model.  
The depth of the circuits described in the previous subsection will depend upon the particular ordering of spin-orbitals chosen for this Jordan-Wigner transform.  We choose to order so that all qubits corresponding to spin up appear first in some order, and then all qubits corresponding to spin down.  Since the hopping terms preserve spin, this simplifies the Jordan-Wigner strings needed.  For a given spin, we order the sites in the ``snake" pattern shown in Fig. \ref{fig:snake}.

With this snake pattern, we group the hopping terms into four different sets, as shown.  The terms in any given set all commute with each other.  The two horizontal sets of terms have no Jordan-Wigner string required, since they involve hopping between neighboring sites with the given ordering pattern, and so these terms can all be executed in parallel.  The vertical terms do not commute with each other.  However, they nest as in Ref.~(\onlinecite{Hastings15}) so that all vertical terms in a given set can be executed with ${\mathcal O}(N)$ gates using a depth ${\mathcal O}(\sqrt{N})$, where we assume that the geometry is roughly square so the the number of sites in a given row is $~\sqrt{N}$.

For periodic boundary conditions, we must also handle the terms going from left to right boundary or from top to bottom boundary.  The terms going from left to right boundary can be executed in parallel with each other.  Each term naively requires a depth proportional to the length of a given horizontal row to calculate the Jordan-Wigner string.  Assuming a square geometry, this length is ${\mathcal O}(\sqrt{N})$.  We can use a tree to calculate the fermionic parity of the sites in the middle of the string, reducing the depth to $\log(N)$ if desired.  For the terms from top to bottom edge, there are Jordan-Wigner strings of length ${\mathcal O}(N)$, but these terms nest and so can all be executed in parallel, reducing the depth to ${\mathcal O}(\log(N))$.  The fermionic parity on the sites on the middle rows (those other than the top and bottom row) can be calculated using a tree, again in depth ${\mathcal O}(\log(N))$.  The total depth can then be ${\mathcal O}(\log(N) \sqrt{N})$.
In fact, using a tree to compute parities, the total depth can be reduced to polylogarithmic in $N$.

Another way to reduce the cost of the Jordan-Wigner transform in this geometry is in Ref.~\onlinecite{VC}.  This method does require additional qubits.

\section{Finding the ground state by quantum phase estimation}
\label{sec:qpe}
The annealing algorithm generates a state with good overlap with the ground state if the rate is sufficiently slow.  \ref{Sec:anneal} discusses optimized paths that increase the overlap.
However, in fact it is not necessary to obtain a perfect overlap with the ground state; if the overlap with the ground state is reasonably large, then we can apply quantum phase estimation to
the state and have a reasonably large chance of projecting onto the ground state.  Since the time for the simplest annealing path scales inversely with the gap squared, while the
time to use quantum phase estimation to measure energies more accurately than the gap (and thus to determine whether or not one is in the ground state) scales inversely with the gap, up to logarithmic corrections, using quantum phase estimation to project an approximate state onto the ground state may be preferred.  The gap will depend upon the annealing path; since we use a symmetry breaking field until nearly the end, the gap only closes at the end of the anneal.  In this section we first give a simple improvement to quantum phase estimation that
leads to a reduction in depth.  We then give some numerical estimates for annealing times for free fermion systems (the simplest case where one can estimate annealing times for a large system numerically).

\subsection{Reducing the number of rotations in quantum phase estimation}
\label{sec:redrot}
While this general setting of using adiabatic preparation and quantum phase estimation has been considered many times before, we briefly note one simplification of the quantum phase estimation procedure.
If the gate set available consists of one- or two-qubit Clifford operations, then this simplification provides a slightly arger than twofold reduction in depth; more importantly, it provides a fourfold reduction in the number
of arbitrary rotations, which are expected to be the most costly operations to implement.  This simplification is not in any way specific to the Hubbard model, and could be applied in other settings, such as in quantum chemistry.
The usual approach to quantum phase estimation is to build controlled unitaries to implement the unitary evolution (by using the ancilla to control the unitaries described above).  In this case, if the ancilla qubit is in the $|0\rangle$ state, then no quantum evolution is implemented, while if the ancilla qubit is in the $|1\rangle$ state, then quantum evolution by $U=\exp(-i H t)$ is implemented (up to Trotter-Suzuki error).
Then, quantum phase estimation estimates the eigenvalues of this unitary $U$.  The simple modification that we propose is that instead, if the ancilla qubit is in the $|0\rangle$ state, then we implement evolution by $\exp(i H t/2)$, while if the ancilla qubit is in the $|1\rangle$ state, then we implement evolution by $\exp(-i H t/2)$.  Since the {\it difference} between these two evolutions is the same $\exp(-i H t)$, then the same phase estimation procedure on the ancilla will return the identical result for the energy (again up to Trotter-Suzuki error).  

This approach means that we have then only to implement unitary evolution for half the time (i.e., $t/2$ rather than $t$), so assuming the same time step for each Trotter step means that the number of Trotter steps required is halved.
However, to analyze whether or not this improves the gate depth, we need to determine how the gate depth might change in a given Trotter step.  First, let us review the usual method of making circuits controlled to implement
phase estimation, and then we will explain how to do it for the modification considered here.  
As an example, let us consider a circuit to implement a hopping term, as in Fig.~\ref{fig:Hpq}.
The usual technique (see references before) to make this circuit controlled (so that if the ancilla is $|0\rangle$ then the identity operation is performed while if the ancilla is $|1\rangle$ then evolution under the hopping term is performed) is to modify
the two $R_z(\theta)$ gates, making them both controlled by an ancilla.
A controlled rotation by angle $\theta$ implements no rotation if the ancilla is in the $|0\rangle$ state and implements a rotation by angle $\theta$ if the ancilla is in the $|1\rangle$ state. 
Different quantum computing architectures may make different elementary gates available.  
 If we have access to arbitrary $Z$ rotations by angle $\theta$, and to Clifford operations, then the controlled rotation by angle $\theta$ is implemented by rotating the target by angle $\theta/2$, then applying a controlled NOT from the ancilla to target, then rotating the target again by angle $-\theta/2$, then again applying a controlled NOT from the ancilla to target, so that the net rotation is $\theta/2-\theta/2=0$ if the ancilla is in the $|0\rangle$ state and is $\theta/2+\theta/2=\theta$ if the ancilla is in the $|1\rangle$ state.  Thus, the controlled rotation is implemented by doing two arbitrary angle rotations, and additionally applying two Clifford operations, for a total of four gates given the gate set mentioned above.

A crucial point is that the {\it only} gates which need to be made controlled are the $R_z(\theta)$ gates.  The other gates in Fig.~\ref{fig:Hpq} do {\it not} need to be modified in any way.  The property still holds when we consider our proposed method of
implementing evolution by either $\exp(i H t/2)$ or $\exp(-i H t/2)$, depending on the ancilla.  In this method, we need to modify the $R_z(\theta)$ gates to implement a rotation by either a positive or negative angle; we term this a ``uniformly controlled rotation": a rotation by angle $+\theta$ if the ancilla is in the $|0\rangle$ state and a rotation by angle $-\theta$ if the ancilla is in the $|1\rangle$ state.  This can be done by
applying a controlled NOT from the ancilla to target, then rotating the target, then again applying a controlled NOT from the ancilla to target.  This requires one arbitary angle rotation and two Clifford operations.

Thus, we find that the total number of gates required to implement the appropriately controlled version of Fig.~\ref{fig:Hpq} has been reduced by $1$, as one fewer arbitrary rotation is required, while the same Clifford operations are required.  So, the depth of a single Trotter step is slightly reduced.
Since the number of Trotter steps has been halved, we find indeed a slightly larger than twofold reduction in depth, as claimed.  Further, the number of arbitrary rotations is reduced by four.  We have described this contruction for the hopping term; however, for any term, the usual technique is to replace arbitrary single qubits rotations by controlled rotations while we suggest instead replacing them with uniformyl controlled rotations.

\subsection{Annealing Time for Free Fermions}

To make some estimate of the cost of adiabatically preparing a good approximation to the ground state, it is worth considering the case of free fermions.  For any system of free fermions with translation symmetry and $k$ sites per unit cell, the annealing problem decouples into many different problems, one for each Fourier mode, each problem describing annealing of a $k$-dimensional subspace.  For $k=2$, this is a Landau-Zener problem.  For a single such Landau-Zener problem, the annealing time is expected to scale as $1/\Delta^2$ where $\Delta$ is the gap of the Hamiltonian.  However, we have many Landau-Zener problems in parallel, and we want all Fourier modes to reach the ground state.  The problem is least serious if the Fermi surface consists just of isolated points (such as in one dimension or for a semimetal in higher dimensions) as there are only a few Fourier modes with small gap.
For systems with a Fermi surface, the density of states at the Fermi surface is larger, meaning that there are more modes with small gap.
However, the probability of a diabatic transition for a Landau-Zener system is exponentially small in the ratio of the gap squared to the Landau-Zener velocity.
Hence, even if there are a large number of modes with gap $\Delta$, this should require only a logarithmic slowdown in the adiabatic velocity to have a small probability of a diabatic transition.
However, one must also worry about effects due to the endpoints of the evolution.  See, however, \ref{Sec:anneal} for a detailed discussion of improved annealing strategies that overcome some of the effects of the endpoints.

To put some numbers onto these generalities, we considered a system of spinless free fermions at half-filling in one dimension with periodic boundary conditions (this of course is the most advantageous case with only two modes with minimum gap).  For each Fourier mode, we compute the evolution using numerically exact techniques (this is a two-dimensional quantum mechanics problem); to determine the probability of the entire system being in the ground state, we multiply the probabilities for each mode.
 We follow the annealing path
\be
H_s=\sum_{i \; {\rm even}} \Bigl( c^\dagger_i c_{i+1}+h.c.\Bigr)+s \sum_{i \; {\rm odd}} \Bigl( c^\dagger_i c_{i+1}+h.c.\Bigr),
\ee
starting at a fully dimerized state at $s=0$ and moving to a uniform state at $s=1$, where $c^\dagger$ ($c_i$) creates (annihilates) a spinless fermion on site $i$ .  We followed a uniform annealing schedule and used a time discretization with negligible Trotter error.
We chose sizes which were not a multiple of $4$ to avoid having an exact zero mode.

For a system $1002$ sites, with an annealing time $t=1000$, the probability of ending in the ground state is $0.54...$.  Thus, the expected time to end in the ground state is roughly $2000$ plus an roughly additional 2 phase estimation steps on average.  Reducing the annealing time to $t=500$, the probability of ending in the ground state is $0.306...$, taking slightly less expected annealing time, but more phase estimation steps on average.  Increasing to $2002$ sites, the situation is worse.  For $t=1000$, the probability of ending in the ground state is $0.13...$, so the expected time is over 7500, plus over 7 phase estimation steps on average.
Because the system is translationally invariant with period $2$ for all $s$, the probability of ending in the ground state is a product over momentum vectors of the probability of ending in the ground state in each momentum vector.  Each of these probabilities can be calculated by evolving a $2$-by-$2$ matrix.
In all cases, the dominant contribution to the smallness of the probability to end in the ground state was from the lowest energy mode.  Thus, small changes in the low energy density of states can lead to dramatic changes in the success probability.

For 1002 sites, the lowest eigenvalue of the free fermion hopping matrix is $0.0188...$, so the gap $\Delta$ of the Hamiltonian is twice that.  Hence, in this case, the phase estimation time could be much smaller than the annealing time.

Since rather large Trotter time steps of order $0.25$ are sufficient to obtain reasonable overlap with the ground state, we find that about  $10^4$ steps are sufficient to prepare the ground state. Using parallel circuit each step can be performed with a small circuit depth that increases only logarithmically with the system size. Hence a parallel circuit depth of about one million gates is sufficient to prepare the ground state. 

\subsection{Improved Annealing Paths}
\label{Sec:anneal}

There are broadly two strategies for reducing the error in the annealing path.  The typical objective in such optimizations is to find a function, $f(s)$, that satisfies $f:[0,1] \mapsto [0,1]$ and $f(0)=0$ and $f(1)=1$ such that the diabatic leakage out of the groundstate caused by $U_{diab}:=e^{iT\int_0^1 E_0(s) \mathrm{d}s}\mathcal{T}e^{-iT\int_0^1 H(f(s)) \mathrm{d} s }$ is minimized for a fixed value of $T$.  Here we use the convention that $\mathcal{T}$ is the time ordering operator, $T$ is the total evolution time allotted and $s=t/T$ is the dimensionless time, and the phase factor is included to ensure that $\lim_{T\rightarrow \infty} U_{diab}=U_{adiab}$ is well defined.

The first strategy is known as \emph{local adiabatic interpolation}.\cite{RC02}  The idea of local adiabatic interpolation is to choose $f(s)$ to increase rapidly when the eigenvalue gap of $H(f(s))$ is large and increase slowly when the gap is small.  This strategy can substantially reduce the time required to achieve a fixed error tolerance, but does not improve the scaling of the diabatic errors in the state preparation, which scale as ${\mathcal O}(1/T)$ for Hamiltonians that are sufficiently smooth.

The second strategy, known as \emph{boundary cancellation} is often diametrically opposed to local adiabatic optimization.\cite{LRH09}  The idea behind it is instead of moving at a speed designed to minimize diabatic transitions, you choose a path designed to maximize cancellations between them.  There are several strategies for exploiting this, but the simplest strategy to to choose $f(s)$ such that $\partial_s^k f(s)|_{s=0,1}=0$ for $k=1,\ldots, m$.  Then if $f(s)$ is at least $m+2$ times differentiable and $H(f(s))$ does not explicitly depend on $T$ then the error in the adiabatic approximation is at most~\cite{WB12}
\begin{equation}
2\left|\max_s\frac{\|\partial_s^{m+1}H(f(s))\|}{\Delta(s)^{m+2}T^{m+1}} \right| + O\left(1/T^{m+2} \right),
\label{eq:adscale}
\end{equation}
where $\Delta(s)$ is the gap at the given value of $s$.
Equation~\eqref{eq:adscale} therefore implies that if
\begin{equation}
f(s) = \frac{\int_0^s y^{m}(1-y)^{m}\mathrm{d}y}{\int_0^1 y^{m}(1-y)^{m}\mathrm{d}y},\label{eq:f}
\end{equation}
then the upper bound on the asymptotic scaling is improved from ${\mathcal O}(1/T)$ to ${\mathcal O}(1/T^{m+1})$.  Furthermore, if a diabatic error of $\delta$ is desired then it suffices to take
\begin{equation}
T= O\left(\max_s\frac{\|\partial_s^{m+1}H(f(s))\|^{\frac{1}{m+1}}}{\Delta(s)^{1+\frac{1}{m+1}}\delta^{\frac{1}{m+1}}} \right).\label{eq:Tscale}
\end{equation}
This implies that even in the limit of small $\delta$ the cost of adiabatic state preparation is not necessarily prohibitive because $m$ can be increased as $\delta$ shrinks to achieve improved scaling with the error tolerance.  Furthermore, the scaling of the error with the minimum gap becomes near--quadratically smaller than would be otherwise expected (for $\delta$ sufficiently small).

The above argument only discusses the scaling of the evolution time under the assumption that $m$ is fixed.  If $m$ grows as $\delta$ shrinks then there may be $m$--dependent prefactors that are ignored in the above analysis.
If the Hamiltonian is analytic in a strip about $[0,1]$ and it remains gapped throughout the evolution then Lidar {\em et al.} show that, for fixed $T$, the error grows at most polynomially with $m$.\cite{LRH09}  Calculus then shows that the optimal value of $m$ scales logarithmically with the error tolerance and hence such contributions are subpolynomial.

These two strategies are often mutually exclusive when the avoided crossing does not occur near the boundary because if $H(s)$ is analytic then $\dot{H}(s)$ will have to be larger away from the boundary to compensate for the fact that it is nearly constant in the vicinity of the boundary.  This can cause the Hamiltonian to move rapidly through the avoided crossing, which increases the annealing time required.  However, the minimum gap often occurs at the end of the evolution when transforming from the initial Hamiltonian (such as the free Fermion model, plaquette preparation, or other tractable approximation) to the interacting theory.  This means that the two strategies are often compatible here and that boundary cancellation methods will often be well suited for high--accuracy state preparation. Further adiabatic optimizations could also be achieved by altering~\eqref{eq:f} to approximate the local--adiabatic path in the region away from from $s=0,1$.

We are of course concerned about the evolution time required to perform $U_{adiab}\approx U_{diab}$ since it dictates the resources needed to prepare the ground state $\psi_0$ within a
fixed error by
adiabatic evolution starting from the free fermion ground
state $\psi_0^ff$.  A related problem is the problem of implementing the projector onto the ground state, $P_0=|\psi_0\rangle\langle \psi_0|$
within a fixed error; we will wish to be able to do this for a method described in \ref{sec:recover}.  Given the projector $P_0^{ff}=|\psi_0^{ff}\rangle\langle \psi_0^{ff}|$ we have $P_0=U_{adiab} P_0^{ff} U_{adiab}^\dagger$; we describe later how to measure $P_0^{ff}$ and so by conjugating this measurement by $U_{adiab}$ we can measure $P_0$.

At first glance, the use of a Trotter decomposition may appear to be problematic for adiabaticity because the time--dependent Hamiltonian that describes the decomposition is discontinuous.  Such discontinuities are not actually problematic, as can be seen by using the triangle inequality
\begin{eqnarray}
&& |(U_{adiab}-U_{diab}^{TS})|\psi_0^{ff}\rangle|  \\ \nonumber
&\leq & \|U_{adiab}-U_{diab}^{TS} \| \\ \nonumber
&\leq & \|U_{adiab} - U_{diab}\| + \|U_{diab} - U^{TS}_{diab}\| ,
\end{eqnarray}
where $U^{TS}_{diab}$ is a quantum circuit approximation to the finite time evolution $U_{\rm diab}$.  
This shows that the error in the state preparation is at most the sum of the adiabatic error and the error in simulating the adiabatic evolution.  
We have a similar bound for error in measuring $P_0$:
\begin{align}
&\|U_{adiab}P_0^{ff}U_{adiab}^{\dagger} - U^{TS}_{diab} P_0^{ff} U^{TS \dagger}_{diab}\|\nonumber\\
&~~\le 
2\left( \|U_{adiab} - U_{diab}\| + \|U_{diab} - U^{TS}_{diab}\| \right).
\end{align}

In order to make the entire simulation error at most $\delta$, it suffices to set both sources of error to $\delta/2$.  This creates an issue because the cost of simulation scales at least linearly with $T$, which in turn scales with $1/\delta$.  There are a wide array of different Trotter--based formulas that can be used in the simulation but in all such cases the Trotter number (and in turn the simulation cost) scales\cite{WBHS10} for integer $k>0$ as $T^{1+1/2k}/\delta^{1/2k}$, which using~\eqref{eq:adscale} is 
$ O\left({\delta^{\frac{-(2k+m+2)}{2k(m+1)}}}\right)$
For any fixed $m$, the value of $k$ that leads to the best scaling with $\delta$ can found by numerically optimization.  However, if we approximate this optimal value by taking $2k=m$ for even $m$ then the cost of the Trotterized simulation is ${\mathcal O}(\delta^{-2/m})$,
which leads to subpolynomial (but not polylogarithmic) scaling of the total cost of simulation in $1/\delta$ if $2k=m$ and the Hamiltonian is sufficiently smooth because the cost of a Trotterized simulation grows exponentially with $k$ unlike the error in the adiabatic approximation.

This shows that the cost of implementing $P_0$ or preparing $\psi_0$ within error $\delta$, $T_0$, is sub--polynomial in $1/\delta$.  Since ${\mathcal O}(\epsilon^{-1}\log(1/\epsilon))$ repetitions of this circuit are needed  and since errors are subadditive it follows that taking $\delta \propto \epsilon^2/\log(1/\epsilon)$ suffices to make the total error at most $\epsilon$.  Thus the cost of implementing $P_0$ scales as $\epsilon^{-1+o(1)}$ where the $o(1)$ term is the amalgamated costs of the adiabatic preparation and the logarithmic term from phase estimation.  

It also should be noted that high--order Trotter--Suzuki formulas may not be needed in practice as numerical evidence suggests that small Trotter numbers typically suffice for the small Hubbard models that we have considered.  This means that low--order methods may often be preferable to their higher--order brethren.  Tight error bounds for the second order Trotter--Suzuki formula are given in the appendix.

\section{Measuring observables}

\label{sec:measurements}
In this section we will discuss the measurement of interesting physical observables. The total energy, which could be measured by phase estimation, is the least interesting quantity. We will instead focus on densities and density correlations, kinetic energies, Green's functions and pair correlation functions.

Such computations will allow us to understand the physics of the ground state. For instance, if the ground state is superconducting,
then we can compute the correlation function of the pair field by the methods described in this section. Its long-distance behavior
determines the condensate fraction. We can also compute the expectation value of the kinetic energy; its variation with doping
can help determine whether the system becomes superconducting as a result of kinetic energy gain compared
to the non-superconducting state. Finally, more complex correlation functions, which would be very difficult -- if not impossible
to determine in experiments -- could determine how these superconducting properties vary in response to perturbations that
enhance or suppress effects such as spin or charge fluctuations.

\subsection{Local observables and equal-time correlations}
\subsubsection{Measuring the density, double occupancy and spin and density correlations}
The densities $n_{i,\sigma}$ are trivially measured by measuring the value of the qubit corresponding to the spin-orbital $(i,\sigma)$. Similarly double occupancies $n_{i,\uparrow}n_{i,\downarrow}$ can be determined by measuring two qubits, while density correlation functions 
\begin{equation}
n_{i}n_{j} = (n_{i,\uparrow}+n_{i,\downarrow}) (n_{j,\uparrow}+n_{j,\downarrow})
\end{equation}
and spin correlation functions
 \begin{equation}
S^z_{i}S^z_{j} = \frac{1}{4}(n_{i,\uparrow}-n_{i,\downarrow}) (n_{j,\uparrow}-n_{j,\downarrow})
\end{equation}
can be determined directly by measuring the values of four qubits. 

By simultaneously measuring all qubits in the computational basis (eigenvectors of the Pauli-$Z$ operator) we can determine density, double occupancy and all spin and density correlations.
As an example, suppose we have recorded this sequence of meaurement outcomes over several runs (for each run, for every qubit we measure the Pauli-$Z$ operator and we record the measurement outcomes).  Given this data, suppose we now wish to
estimate a correlation function such as $$\langle \psi_0 | n_{i,\uparrow} n_{j,\uparrow} | \psi_0 \rangle$$ for a pair of sites $i \neq j$.  This is equal to the expectation value of $$\frac{Z_{i,\uparrow} +1}{2}\frac{Z_{j,\uparrow}+1}{2}.$$  Given the outcomes of the measurements of $Z_{i,\uparrow},Z_{j,\uparrow}$, we simply add $1$ to each measurement, multiply the results, and divide by four.  We then average this over runs.  For example, if we consider a two-site Hubbard model at half-filling and $U=0$, we will find that the two measurement outcomes $Z_{i,\uparrow}$ and $Z_{j,\uparrow}$ are perfectly anti-correlated (since there is only one electron with spin up which has equal probability to be on either of the two sites) and so the average of the product will not equal the product of the averages.

 \subsubsection{General strategy for other observables}
\begin{figure}
\begin{tikzpicture}[scale=1.00,every node/.style={scale=1.00}]
\matrix[matStyle] {
  \lbl[0-0]{\ket{0}} & \point[0-1] &  & \gate[0-3]{H} & \gate[0-4]{Z(\theta)} & \ctrl[0-5] & \gate[0-6]{H} & \meter[0-7] & \point[0-8]  \\
  \point[1-0] & \point[1-1] &  &  &  & \gate[1-5]{U} &  &  & \point[1-8]  \\
   &  &  &  &  &  &  &  &   \\
};
\begin{pgfonlayer}{background}
\qw{0-1}{0-7} \qw{1-1}{1-8} \cw{0-7}{0-8} 
\qwx{0-5}{1-5} 
\end{pgfonlayer}
\end{tikzpicture}
\caption{\label{fig:PE} General phase estimation circuit to compute the expectation value of any unitary which can be given as a controlled black box. The $Z(\theta)$ produces a rotation about the $Z$ axis by angle $\theta$; by varying this, real and imaginary parts of the expectation value can be measured.
}
\end{figure}
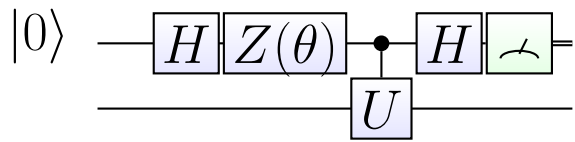

Next, we note that there is a {\it general} strategy used to measure the expectation value of any unitary operator $U$, assuming that we can build a circuit that implements a controlled version of this unitary, controlled by some ancilla.
Namely, we apply a one-bit phase estimation using the phase estimation circuit of Fig. \ref{fig:PE}.
This is a standard trick; see for example Fig.~9 in Ref. \onlinecite{Somma2002}.
Since we have circuits that implement unitary evolution under various Hamiltonian terms, this enables us to meaure these terms.  For example, to measure a term
$c^\dagger_{p,\sigma} c_{q,\sigma}$, we measure the expectation value of the unitary $\exp[-i \theta (c^\dagger_{p,\sigma} c_{q,\sigma}+h.c.)]$.  Since the operator $c^\dagger_{p,\sigma} c_{q,\sigma}+h.c.$ has eigenvalues $-1,0,+1$, the most efficient results are obtained from the phase estimation when we choose $\theta=\pi$ (which perfectly distinguishes in a single measurement between eigenvalue $0$ and eigenvalues $\pm 1$) or $\theta=\pi/2$ (which perfectly distinguishes the case of eigenvalue $+1$ from the case of eigenvalue $-1$).

However, for the observables we consider, there in fact are much simpler ways of measuring the correlation functions.  We give two different strategies.  The first strategy involves replacing rotations in some of our unitary gates with measurements and we call it the ``stabilizer strategy"; the second introduces a new gate called an ``FSWAP".

\subsubsection{The Stabilizer Strategy}
The stabilizer strategy is a method for measuring observables of the form $\exp[-i\theta (O_1 + O_2 + ...)]$ or of the form $O_1+O_2+...$, where each $O_i$ is a product of some number of Pauli operators, and $[O_i,O_j]=0$.  
This form includes many operators of interest to us, including terms in the Hamiltonian such as the kinetic energy as well as other terms such as pair correlation.
We call this the stabilizer strategy because of our use of products of Paulis which commute with each other; no assumption is made that any state is a stabilizer state.  

\begin{figure*}
\begin{tikzpicture}[scale=0.70,every node/.style={scale=0.70}]
\matrix[matStyle] {
  \point[0-0] & \gate[0-1]{H} &  & \ctrl[0-3] &  &  &  &  &  &  &  &  &  &  & \ctrl[0-14] &  & \gate[0-16]{H} & \gate[0-17]{Y} &  & \ctrl[0-19] &  &  &  &  &  &  &  &  &  &  & \ctrl[0-30] &  & \gate[0-32]{Y^\dagger} & \point[0-33]  \\
  \point[1-0] &  &  & \targ[1-3] & \ctrl[1-4] &  &  &  &  &  &  &  &  & \ctrl[1-13] & \targ[1-14] &  &  &  &  & \targ[1-19] & \ctrl[1-20] &  &  &  &  &  &  &  &  & \ctrl[1-29] & \targ[1-30] &  &  & \point[1-33]  \\
  \point[2-0] &  &  &  & \targ[2-4] & \ctrl[2-5] &  &  &  &  &  &  & \ctrl[2-12] & \targ[2-13] &  &  &  &  &  &  & \targ[2-20] & \ctrl[2-21] &  &  &  &  &  &  & \ctrl[2-28] & \targ[2-29] &  &  &  & \point[2-33]  \\
  \point[3-0] &  &  &  &  & \targ[3-5] & \ctrl[3-6] &  &  &  &  & \ctrl[3-11] & \targ[3-12] &  &  &  &  &  &  &  &  & \targ[3-21] & \ctrl[3-22] &  &  &  &  & \ctrl[3-27] & \targ[3-28] &  &  &  &  & \point[3-33]  \\
  \point[4-0] &  &  &  &  &  & \targ[4-6] & \ctrl[4-7] &  &  & \ctrl[4-10] & \targ[4-11] &  &  &  &  &  &  &  &  &  &  & \targ[4-22] & \ctrl[4-23] &  &  & \ctrl[4-26] & \targ[4-27] &  &  &  &  &  & \point[4-33]  \\
  \point[5-0] &  & \gate[5-2]{H} &  &  &  &  & \targ[5-7] & \meter[5-8] & \gate[5-9]{\ket{M}} & \targ[5-10] &  &  &  &  & \gate[5-15]{H} &  &  & \gate[5-18]{Y} &  &  &  &  & \targ[5-23] & \meter[5-24] & \gate[5-25]{\ket{M}} & \targ[5-26] &  &  &  &  & \gate[5-31]{Y^\dagger} &  & \point[5-33]  \\
   &  &  &  &  &  &  &  &  &  &  &  &  &  &  &  &  &  &  &  &  &  &  &  &  &  &  &  &  &  &  &  &  &   \\
};
\begin{pgfonlayer}{background}
\qw{0-0}{0-33} \qw{1-0}{1-33} \qw{2-0}{2-33} \qw{3-0}{3-33} \qw{4-0}{4-33} \qw{5-0}{5-8} 
\qw{5-9}{5-24} \qw{5-25}{5-33} \cw{5-8}{5-9} \cw{5-24}{5-25} 
\qwx{0-3}{1-3} \qwx{0-14}{1-14} \qwx{0-19}{1-19} \qwx{0-30}{1-30} \qwx{1-4}{2-4} 
\qwx{1-13}{2-13} \qwx{1-20}{2-20} \qwx{1-29}{2-29} \qwx{2-5}{3-5} \qwx{2-12}{3-12} 
\qwx{2-21}{3-21} \qwx{2-28}{3-28} \qwx{3-6}{4-6} \qwx{3-11}{4-11} \qwx{3-22}{4-22} 
\qwx{3-27}{4-27} \qwx{4-7}{5-7} \qwx{4-10}{5-10} \qwx{4-23}{5-23} \qwx{4-26}{5-26} 

\end{pgfonlayer}
\end{tikzpicture}
\caption{\label{fig:HpqMeas} $H_{pq}$ Measurement}
\end{figure*}
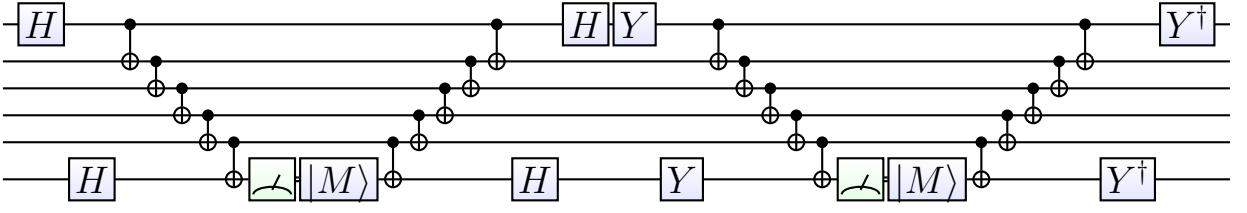

If we can measure each $O_i$, then we succeed in measuring the desired operator.  Since they commute, we may measure them in any order without needing to recreate the state after measurement.  As each $O_i$ is a product of Paulis, there is some unitary in the Clifford group which maps it onto a Pauli $Z$ operator on some given qubit.  Hence, we can measure
that $O_i$ by applying that Clifford unitary, then doing a $Z$-basis measurement, and finally undoing the Clifford unitary.
Applying this procedure to terms in the Hamiltonian, for which we have previously given circuits to implement $\exp[-i \theta(O_1+O_2+...)]$ using sequences of controlled $Z$-basis rotations conjugated by Clifford gates, the measurement circuit amounts to replacing each controlled $Z$-basis rotation in the evolution circuit with a $Z$-basis measurement.  For example, if we apply this procedure to the $H_{pq}$ unitary evolution shown in Fig.~\ref{fig:Hpq}, we arrive at the circuit shown in Fig.~\ref{fig:HpqMeas}.  Summing the value of the two $Z$-measurements (treating the answers as $\pm 1$ for each measurement) and then dividing by two (this division occurs because the controlled unitaries rotate by an angle equal to half the coupling strength) gives a measurement of the expectation operator of the hopping operator. 
In fact, we find that the {\it same} measurements are needed to measure the pairing operator for $\Delta$ real, since as we have noted, the pairing operator for $\Delta$ real is implemented by the same unitary as the hopping operator, up to a sign change in the second controlled rotation.  Thus, the two measurements we have given {\it simultaneously} measure the two commuting operators $c^\dagger_{p,\sigma} c_{q,\sigma'}+h.c.$ and $c^\dagger_{p,\sigma} c^\dagger_{q,\sigma'}+h.c.$ (to measure the pairing operator, one must instead consider the difference of the two $Z$ measurements).

\subsubsection{The FSWAP Strategy for the kinetic energy and Green's functions}
For the kinetic energies we will have to measure one-body  Green's functions $c^\dag_{i,\sigma}c_{j,\sigma}+c^\dag_{j,\sigma}c_{i,\sigma}$. To do this we first swap qubit $j$ with its neighboring qubits until it is neighboring qubit $i$ in the normal ordering and is at one of the positions $k=i\pm1$. Taking into account the fermionic nature of the electrons we need to use a  new {\it fermionic swap gate} FSWAP with matrix
\begin{equation}
\left( \begin{array}{cccc}1&0&0&0 \\ 0 & 0 & 1 & 0 \\  0 & 1 & 0 & 0 \\ 0 & 0 & 0 & -1 \end{array}\right).
\end{equation}
The circuit expressed in standard gates is:

\begin{equation}
\begin{tikzpicture}[scale=1.00,every node/.style={scale=1.00}]
\matrix[matStyle] {
  \point[0-0] & \qswap[0-1] & & & \qswap[0-2] &  & \ctrl[0-4] &  & \point[0-6]  \\
   &  & \lbl{=}  & &  &  &   \\
  \point[1-0] & \qswap[1-1] & & & \qswap[1-2] & \gate[1-3]{H} & \targ[1-4] & \gate[1-5]{H} & \point[1-6]  \\
   &  &  &  &  &  &   \\
};
\begin{pgfonlayer}{background}
\qw{0-2}{0-6} 
\qw{1-2}{1-6} 
\qwx{0-2}{1-2} \qwx{0-4}{1-4} \fwx{0-1}{1-1} 
\end{pgfonlayer}
\end{tikzpicture}
\end{equation}

We next change the basis to be able to better measure the kinetic energy on the bond which is then $-t_{pq}\left(c^\dag_{p,\sigma}c_{q,\sigma} + c^\dag_{q,\sigma}c_{p,\sigma} \right)$. This basis change is a simple circuit:

\begin{equation}
\begin{tikzpicture}[scale=1.0,every node/.style={scale=1.0}]
\matrix[matStyle] {
  \point[0-0] &  \targ[0-3] &  & \point[0-5]  \\
  \point[1-0] &  \ctrl[1-3] & \gate[1-4]{H} & \point[1-5]  \\
};
\begin{pgfonlayer}{background}
\qw{0-0}{0-5} \qw{1-0}{1-5} 
\qwx{0-3}{1-3}

\end{pgfonlayer}
\end{tikzpicture}
\end{equation}

giving the  unitary:

\begin{equation}
\left( \begin{array}{cccc}1/\sqrt{2} &0&0& 1/\sqrt{2}  \\ -1/\sqrt{2}  & 0 & 0 & 1/\sqrt{2}  \\  0 & 1/\sqrt{2} & 1/\sqrt{2} & 0 \\ 0 & -1/\sqrt{2} & 1/\sqrt{2} & 0 \end{array}\right).
\end{equation}

\begin{figure}
\begin{tikzpicture}[scale=1.00,every node/.style={scale=1.00}]
\matrix[matStyle] {
  \point[0-0] &                        &                    & \targ[0-3] &  & \meter[0-5] & \point[0-6]  \\
  \point[1-0] &                        & \qswap[1-2] & \ctrl[1-3] & \gate[1-4]{H} & \meter[1-5] & \point[1-6]  \\
  \point[2-0] & \qswap[2-1]     & \qswap[2-2] & \targ[2-3] &  & \meter[2-5] & \point[2-6]  \\
  \point[3-0] & \qswap[3-1]    &                    & \ctrl[3-3] & \gate[3-4]{H} & \meter[3-5] & \point[3-6]  \\
  \point[4-0] &                       &                    & \targ[4-3] &  & \meter[4-5] & \point[4-6]  \\
  \point[5-0] &                       & \qswap[5-2] & \ctrl[5-3] & \gate[5-4]{H} & \meter[5-5] & \point[5-6]  \\
  \point[6-0] & \qswap[6-1] & \qswap[6-2] & \targ[6-3] &  & \meter[6-5] & \point[6-6]  \\
  \point[7-0] & \qswap[7-1] &  & \ctrl[7-3] & \gate[7-4]{H} & \meter[7-5] & \point[7-6]  \\
   &  &  &  &  &  &   \\
};
\begin{pgfonlayer}{background}
\qw{0-0}{0-5} \qw{1-0}{1-5} \qw{2-0}{2-5} \qw{3-0}{3-5} \qw{4-0}{4-5} \qw{5-0}{5-5} 
\qw{6-0}{6-5} \qw{7-0}{7-5} \cw{0-5}{0-6} \cw{1-5}{1-6} \cw{2-5}{2-6} \cw{3-5}{3-6} 
\cw{4-5}{4-6} \cw{5-5}{5-6} \cw{6-5}{6-6} \cw{7-5}{7-6} 
\qwx{0-3}{1-3} \qwx{2-3}{3-3} 
\qwx{4-3}{5-3} \qwx{6-3}{7-3} \fwx{1-2}{2-2} \fwx{5-2}{6-2} \fwx{2-1}{3-1} \fwx{6-1}{7-1} 

\end{pgfonlayer}
\end{tikzpicture}
\caption{\label{fig:HubGreen-1} Example of a circuit for parallel measurements of a number of Green's function values between different sizes}
\end{figure}
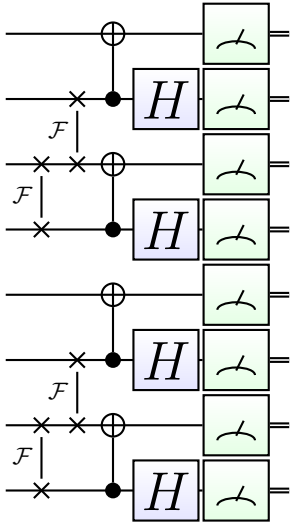

An example circuit containing several measurements in parallel is shown in Fig. \ref{fig:HubGreen-1}. 

We may then measure the $Z$ eigenvalues of the two qubits and call them $s_1$ and $s_2$.  If we measure $s_1=-1$, there is one particle on the two sites. We then have as estimators for the Green's function  $c^\dag_{p,\sigma}c_{q,\sigma'} + c^\dag_{q,\sigma'}c_{p,\sigma}$:
\begin{equation}
s_2\delta_{s_1,-1}
\end{equation}
and for the kinetic energy $-t_{pq}\left(c^\dag_{p,\sigma}c_{q,\sigma} + c^\dag_{q,\sigma}c_{p,\sigma} \right)$ the estimator 
\begin{equation}
- t_{pq}s_2\delta_{s_1,-1}
\end{equation}
Note that if we have $N$ qubits we can do $N/2$ measurements simultaneously, as long as no qubit is involved in two measurements.

Conversely, if the first qubit is +1, there are either 0 or 2 particles on the bond. In the presence of a pairing term, which breaks particle number conservation we can measure non-trivial values for the pair field operator $c^\dagger_{p,\sigma} c^\dagger_{q,\sigma'}+h.c.$ as $s_2\delta_{s_1,1}$.

 \subsubsection{The FSWAP Strategy for the pair correlation functions}
 Pair correlation functions in their most general form are built from the general 2-body terms
$c^\dag_{i,\sigma}c^\dag_{j,-\sigma}c_{k,-\sigma'}c_{l,\sigma}$. This expectation value can be measured after fermion-swapping the qubits to be adjacent and then performing two Green's function measurements. We first use FSWAP gates to move the qubits $i$, $k$, $j$, and $l$ to four adjacent positions which we will label 1 to 4. We then apply the same circuit as for the Green's function measurements to both pairs $(1,2)$ and $(3,4)$ and measure the $Z$ component of each qubit $m$ (=1,2,3, or 4) as $s_m$. The expectation value is then expressed as
$\langle c^\dag_{i,\sigma}c_{k,\sigma} c^\dag_{j,\sigma'}c_{l,\sigma'} \rangle = \langle s_2s_4\delta_{s_1,-1}\delta_{s_3,-1}  \rangle$

Care must be taken if two indices are the same, e.g. if $i=k$ in above term. In that case things simplify since 
\begin{equation}
\langle c^\dag_{i,\uparrow}c_{i,\uparrow} c^\dag_{j,\downarrow}c_{l,\downarrow} \rangle = \langle n_{i,\uparrow}c^\dag_{j,\downarrow}c_{l,\downarrow} \rangle 
\end{equation}
and the measurement just becomes
\begin{equation}
  \langle n_{i,\uparrow}c^\dag_{j,\downarrow}c_{l,\downarrow} \rangle = \langle n_{i,\uparrow} s_4\delta_{s_3,-1}  \rangle
\end{equation}
Similarly if $i=k$ and $j=l$ we get
\begin{equation}
\langle c^\dag_{i,\uparrow}c_{i,\uparrow} c^\dag_{j,\downarrow}c_{j,\downarrow} \rangle = \langle n_{i,\uparrow}n_{j,\downarrow}\rangle 
\end{equation}
and the measurement  is straightforward.

For the Hubbard model we are interested in singlet pairing and the specific terms we want are:
\begin{equation}
\Delta_{ij,kl} = \frac{1}{2}\left(c^\dag_{i,\uparrow}c^\dag_{j,\downarrow} - c^\dag_{i,\downarrow}c^\dag_{j,\uparrow}\right)\left(c_{k,\uparrow}c_{l,\downarrow} - c_{k,\downarrow}c_{l,\uparrow}\right).
\end{equation}
Multiplying this out and sorting by spin we get
\begin{eqnarray}
\Delta_{ij,kl} &=& -\frac{1}{2}\left(c^\dag_{i,\uparrow}c_{k,\uparrow} c^\dag_{j,\downarrow}c_{l,\downarrow} + c^\dag_{i,\uparrow}c_{l,\uparrow} c^\dag_{j,\downarrow}c_{k,\downarrow}  \right. \nonumber \\ && \left. + c^\dag_{j,\uparrow}c_{k,\uparrow} c^\dag_{i,\downarrow}c_{l,\downarrow} + c^\dag_{j,\uparrow}c_{l,\uparrow} c^\dag_{i,\downarrow}c_{k,\downarrow}   \right),
\end{eqnarray}
which can be measured by individually measuring all terms. 

Note that even though there may be $N^4$ different choices for the four indices, we do not need to measure all of these four-point functions. As pairs are expected to be tightly bound in the Hubbard model, choosing $i$ close to $j$ and $k$ close to $l$, but choosing the pairs $(i,j)$ and $(k,l)$ at as large a distance as possible on a given lattice is sufficient to check for long-range pair correlations. 
In fact, we can measure $c^\dag_{i,\uparrow}c^\dag_{j,\downarrow}+ h.c$ (or, $c^\dag_{i,\uparrow}c^\dag_{j,\downarrow}- h.c$) for
$N/2$ pairs $(i,j)$ simultaneously because the operators commute; then, averaging the result over pairs extracts the long wavelength pair correlation.  To measure these with minimum depth, one can use nesting strategies similar to those discussed for measuing hopping terms in Sec.~\ref{optcost}.

 \subsection{Dynamic correlation functions and gaps}

 \subsubsection{Dynamic response in the time domain}

The measurement of dynamic correlation functions
 \begin{eqnarray}
 C_{A,B}(t) &=& \langle \psi_0 | A (t) B(0) | \psi_0 \rangle \\ 
  &\equiv& \langle \psi_0 | e^{+itH} A^\dag  e^{-itH} B  | \psi_0\rangle \nonumber
 \end{eqnarray}
for time-independent Hamiltonians such as the Hubbard model are typically presented not in the time domain but after a Fourier transform in the frequency domain:
 \begin{eqnarray}
 S_{A,B}(\omega) &=& \int_{-\infty}^\infty dt  C_{A,B}(t) e^{i\omega t}
  \end{eqnarray}
The standard way of measuring them has been explained in several references, most explicitly in Ref. \onlinecite{Somma2002}. For unitary operators $A,B$, the product $A^\dagger e^{+itH} B e^{-itH}$ is a unitary, and can be measured using the
circuit of Fig.~\ref{fig:PE}.  
One simple improvement is to note that it is not necessary to control the evolutions $e^{itH},e^{-itH}$ and it suffices to control $A^\dagger,B$; then, if the ancilla bit is $|0\rangle$ so that the operators $A^\dagger,B$ are not done, the product $e^{itH} e^{-itH}$ is equal to the identity as desired.  The uncontrolled time evolution will require only half as many arbitrary angle rotations as the controlled time evolution for many gate sets (see Sec.~\ref{sec:redrot} for discussion
of implementing controlled time evolution).

A further improvement is to note that the final evolution $e^{-itH}$ just gives an overall phase acting on the
ground state, and so can be replaced by $Z$ rotation of the ancilla without being implemented, giving another factor of two reduction in depth.
A final improvement combines the idea in Ref.~\onlinecite{Somma2002} of controlling one operation and anti-controlling another with one of the
ideas in Sec.~\ref{sec:redrot} of implementing either forward or backward evolution in time.  We prepare the ancilla in the state $|+\rangle$.  If the ancilla is $|0\rangle$, we implement
$e^{itH/2} B$ and if the ancilla is $|1\rangle$ we implement $e^{-itH/2} A$.  The measurement of the ancilla in the $X$ basis gives the desired
expectation value.  The controlled time evolution, by either $e^{itH/2}$ or $e^{-itH/2}$ depending on whether the ancilla is $|0\rangle$ or $|1\rangle$,
can be done as described in Sec.~\ref{sec:redrot}.  This requires half the number of time steps as that required implement the uncontrolled evolution by $e^{itH}$, assuming that the same timestep is used for both evolutions; the only additional cost is an additional CNOT gate for each arbitrary angle $Z$ rotation in the evolution and since these are Clifford gates the cost is much smaller than the cost of an arbitrary angle rotation for many gate sets.

 \subsubsection{Dynamic response in the frequency domain by phase estimation}
 
A dynamic correlation function of an operator with its adjoint, such as
 \begin{eqnarray}
 C_{A}(t) &=& \langle \psi_0 | A^\dag (t) A(0) | \psi_0 \rangle \\ 
  &\equiv& \langle \psi_0 | e^{+itH} A^\dag e^{-itH} A   | \psi_0\rangle, \nonumber
 \end{eqnarray}
can be simplified after a Fourier transform into the frequency domain:
 \begin{eqnarray}
 S_{A}(\omega) &=& \int_{-\infty}^\infty dt  \, e^{i\omega t}\, C_{A}(t)  \\
 &=&  \int_{-\infty}^\infty dt \,e^{i\omega t}  \langle \psi_0 | e^{+itH} A^\dag e^{-itH} A   | \psi_0\rangle \nonumber \\
 &=&  \int_{-\infty}^\infty dt \,e^{i\omega t}  \sum_n \langle \psi_0 |  e^{+itH} A^\dag e^{-itH/\hbar}
|{\psi_n}\rangle \langle\psi_{n}| A | \psi_0\rangle \nonumber \\
 &=&  \int_{-\infty}^\infty dt \sum_n e^{i(\omega-(E_n-E_0)) t}  \langle \psi_0 | A^\dag|{\psi_n}\rangle  \langle{\psi_n} | A | \psi_0\rangle \nonumber \\
 &=&  \sum_n  |\langle {\psi_n} | A | \psi_0\rangle|^2 \nonumber \delta(\omega-(E_n-E_0))  .
 \end{eqnarray}

Instead of performing a ``simple sampling'' and measuring $ C_{A}(t) $ for all times $t$ and then Fourier-transforming it and hoping to get these delta-functions resolved a new and better approach is to do ``importance sampling'' and measure the energies of the eigenstates $|n\rangle$ with eigenenergies $E_n$ directly with the weight $|\langle {\psi_n} | A | \psi_0\rangle|^2$  with which they appear in above sum. 

This can be achieved by phase estimation. If we apply phase estimation to the state $A | \psi_0\rangle$ then eigenstate $|{\psi_n}\rangle$ will be picked just with the weight A $|\langle {\psi_n} | A | \psi_0\rangle|^2$ and a histogram of the measured energies thus directly measures the dynamic structure factor $ S_{A}(\omega)$.

The retarded correlation function:
 \begin{equation}
 \chi_{A}(t) = \theta(t) \langle \psi_0 | \left[ A^\dag (t), A(0)\right] | \psi_0 \rangle 
 \end{equation}
can then be obtained as follows. First, we write the retarded correlation function in terms of
its real and imaginary parts, $\chi_{A}(\omega) = \chi'_{A}(\omega) + i \chi''_{A}(\omega)$.
The imaginary part of the retarded correlation function, $\chi''_{A}(\omega)$ is then obtained from
the fluctuation-dissipation theorem: $\chi''_{A}(\omega) = (1-e^{-\beta \omega}) S_{A}(\omega)$;
at zero-temperature, $\beta=\infty$, and the second term in parentheses vanishes. 
The real part $\chi'_{A}(\omega)$
is then obtained by the Kramers-Kronig relation:
\begin{equation}
\chi'_{A}(\omega) = {\cal P} \int_{-\infty}^{\infty} \frac{d\omega'}{\pi} \, \frac{\chi''_{A}(\omega)}{\omega'-\omega}
\end{equation}

By choosing $A=Z_{p,\sigma}$ we can measure the local dynamical spin density correlations. Explicitly we find that
\begin{eqnarray}
\langle Z_{p,\sigma}(0) Z_{p',\sigma'}(t) \rangle  &=& 4 \langle n_{p,\sigma}(0) n_{p',\sigma'}(t)  \rangle \\
&&- 2 n_{p,\sigma}(0) -2 n_{p',\sigma'}(t)  +1. \nonumber
\end{eqnarray}
Since $n_{p',\sigma'}(t) = n_{p',\sigma'}(0)$ the last three terms do not depend on $t$ and only contribute a constant, which does not give any contribution to the interesting finite-$\omega$ structure factor. Local dynamical charge and spin correlations are straightforwardly calculated from the spin density correlations.

Similarly, local single-particle excitations can be measured by choosing 
\begin{equation}
A=c_{p,\sigma}^\dagger+c_{p,\sigma} = \prod_{q<p}Z_{q,\sigma}X_{q,\sigma}.
\end{equation}
Whether a particle or a hole excitation is realized in the excited state that is picked out by phase estimation can easily be determined by measuring the total particle number of that state.

Using circuits similar to the preparation of Slater determinants, we can measure dynamical correlation functions not only locally but for arbitrary single-particle wave functions. In particular, using momentum eigenstates one can obtain the momentum-resolved electron and hole spectral functions as measured in angle resolved photo emission (ARPES).

\section{Non-Destructive Measurements}

\label{sec:non-destructive}

For the small systems that we have explicitly simulated, the time needed to adiabatically prepare a good estimate
of the ground state is small compared to the phase estimation time.
However, this will change as one increases the system size, and the time to adiabatically prepare the ground state will ultimately dominate, at least if one uses an
annealing path that linearly interpolates between initial and final Hamiltonian.
The reason is the time to adiabatically prepare the ground state is expected to scale as $\Delta^{-2}$  for a system with spectral gap $\Delta$, while resolving energies to accuracy $\Delta$ with phase estimation takes time $1/\Delta$ and thus for small $\Delta$ the phase estimation is faster.  
We note that the improved higher-order annealing paths in \ref{Sec:anneal} do alleviate this problem.

This problem is slightly alleviated by the fact that we are content to prepare by annealing a state with significant overlap (say, $1/2$ overlap) with the ground state, as then
phase estimation will give us a significant chance of projecting onto the ground state.  However, once we have
prepared a ground state $\psi_0$ by a combination of annealing and phase estimation, it will be desirable to measure properties of the state without 
destroying it. Here we present two approaches for such non-destructive measurements.  Section \ref{sec:hm} discusses an approach based on the Hellman-Feynman theorem, which not only is non-destructive but also scales better in the error than the simple way of repeatedly preparing the ground state and measuring . An alternate approach based on ``recovering" the state after measurements may be more useful in some circumstances and is discussed in Sec. \ref{sec:recover}.

 \subsection{Hellman-Feynman based approach}
 \label{sec:hm}
 
According to the Hellman Feynman theorem  the derivative of the ground state energy $E_0(\lambda)$ with respect to a perturbation $\lambda  { O}$ is just the expectation value of the operator $O$  in the ground state $|\Psi_0(\lambda)\rangle$:
 \begin{eqnarray}
\frac{d}{d\lambda} E_0(\lambda)& =& \frac{d}{d\lambda} \langle \Psi_0(\lambda)| H+\lambda { O} | \Psi_0(\lambda) \rangle \nonumber \\ &=&  \langle \Psi_0(\lambda)|  { O} | \Psi_0(\lambda) \rangle.
 \end{eqnarray}
 
  An alternative and superior way of measuring expectation values of arbitrary observables ${ O}$ is to adiabatically add a small perturbation $\lambda { O}$ to the Hamiltonian $H$.

 This opens a way to measuring observables without destroying the ground state wave function $ | \Psi_0(0) \rangle $  of $H$. We adiabatically evolve the ground state wave function $ | \Psi_0(0) \rangle $ to $ | \Psi_0(\lambda) \rangle $ by slowly increasing $\lambda$ to its final (small) value, and then perform another quantum phase estimation to determine the energy $E_0(\lambda)$. The ground state expectation value 
can be estimated through
 \begin{equation}
 \langle \Psi_0(0) |  { O} | \Psi_0(0) \rangle = \frac{d}{d\lambda} E_0(\lambda)|_{\lambda =0} = \frac{E_0(\lambda)-E_0(0)}{\lambda} + {\mathcal O}(\lambda^2).
 \end{equation}
 For example, to measure a Green's function we add a ``hopping'' term $\lambda(c^\dag_{p,\sigma}c_{q,\sigma}+c^\dag_{q,\sigma}c_{p,\sigma})$ to the Hamiltonian. 
 
 In order to obtain an estimate with an error $\epsilon$ we need to choose $\lambda ={\mathcal O}(\epsilon)$, and then measure the energy to an accuracy $\lambda\epsilon=
 {\mathcal O}(\epsilon^2)$. Phase estimation then requires time  ${\mathcal O}(\epsilon^{-2})$, which scales the same as repeated preparation with destructive measurements and thus only gets a constant improvement.
 
 The scaling can be improved by using a symmetric estimator for the derivative
  \begin{equation}
 \langle \Psi_0(0) |  { O} | \Psi_0(0) \rangle =  \frac{E_0(\lambda)-E_0(-\lambda)}{2\lambda} + {\mathcal O}(\lambda^3),
 \end{equation}
 which reduces the approximation error. To get the energy  and thus allows a larger value of $\lambda ={\mathcal O}(\epsilon^{1/2})$, which requires phase estimation only to an accuracy of  ${\mathcal O}(\epsilon^{3/2})$, and a time scaling as ${\mathcal O}(\epsilon^{-3/2})$. In general, using a $k$-th order estimator for the derivative requires ${\mathcal O}(k)$ energy measurements, but with an error scaling as ${\mathcal O}(\lambda^k)$, we can choose $\lambda ={\mathcal O}(\epsilon^{1/(k-1)})$, and require time ${\mathcal O}(k\epsilon^{-1-1/(k-1)})$. Asymptotically for small $\epsilon$ we can thus estimate the expectation value non-destructively with an effort scaling as ${\mathcal O}(-\frac{\log\epsilon}{\epsilon})$, obtaining a near-quadratic speedup.
 
 In the Hellman-Feynman approach, there may be no need to do an adiabatic evolution.  In many cases we could simply start in the state $\psi_0$, then apply a phase estimation to expectation value of $H+\lambda O$, and then apply another phase estimation to estimate the ground state energy of $H$.  The absolute value squared of the overlap between the ground state of $H$ and the ground state of $H+\lambda O$ is equal to $1-{\mathcal O}(\lambda^2/\Delta^2)$, and so
for $\lambda<<\Delta$ the overlap is close to unity.  So, with probability close to $1$, the energy given by phase estimation of $H+\lambda O$ will in fact be the desired ground state energy, and with again probability close to $1$ the second phase estimation will return to the ground state of $H$.

\subsection{Non-Destructive Measurements With Quadratic Speedup}
\label{sec:recover}

This annealing procedure becomes less useful for operators $O$ which require a complicated quantum circuit.  Suppose we wish to measure the expectation value of a time dependent correlation such as $O=S^z(t) S^z(0)$, where $t>0$ and where $S^z(t)=\exp(i H t)S^z \exp(-iHt)$.  In this case, the quantum circuit to measure $O$ (given below) is much more complicated; thus the Hamiltonian $H+\epsilon O$ becomes significantly more costly to do phase estimation on.

In this section, we explain a technique to measure arbitrary projection operators in a non-destructive fashion.  We then further explain a quadratic speedup of this approach.  Our initial approach will give many independent binary measurements with outcome probability determined by the expection value of the operator. This will require a number of samples proportional to the square of the inverse error. On the other hand, the improved approach will exploit phase estimation to speed this up quadratically (up to log factors).

As we have explained previously, we have quantum circuits that implement projective measurements of all the operators we are interested in, such as number operator, hopping operator, and so on.  Further, we have a general technique for implementing a projective measurement of any unitary that can be controlled.

\subsubsection{Recovery Map}
Suppose we wish to implement a projective measurement with $k$ possible outcomes, for some given $k$.  We describe these outcomes by projectors $Q_1,...,Q_k$, with $\sum_i Q_i=I$, and we refer to this measurement as ``measuring Q".  The ground state $\psi_0$ can be written as a linear combination of states in the range of each of these projectors:
\be
\psi_0=\sum_i a_i \phi_i,
\ee
where
\be
Q_i \phi_i = \phi_i,
\ee
and $|\phi_i|=1$.
Assume for the moment that we can also implement a measurement $P_0$ which projects onto the ground state $\psi_0$ (we describe how to do this to sufficient accuracy below).
Then, the algorithm for nondestructive measurement is: first, begin in the ground state.  Then, measure $Q$.  Then, measure $P_0$.  If the result is that one indeed is in the ground state, then we have restored the ground state; at this point we can either re-measure the projector $Q$ to obtain better statistics, or make some other measurement of a different observable.  If instead the measurement of $P_0$ reveals that we are not in the ground state, we re-measure $Q$ and then re-measure $P_0$.  We repeat this process of re-measuring $Q$ and re-measuring $P_0$ until we are in the ground state.  The basic idea is similar to that in Ref.~\onlinecite{PV}.

The rest of this subsubsection is focused on calculating the probability of returning to the ground state; most readers may wish to skip to the next subsubsection, which gives the more important quadratic speedup.  The main reason for introducing the recovery map in this subsubsection is that it can be used after the quadratic speedup (see below).
Let us compute the probability of returning to the ground state in this process.  If the initial measurement of $Q$ gives outcome $i$ then the resulting state is $\phi_i$; since, $|\langle \psi_0 | \phi_i \rangle|^2=|a_i|^2$, we have a probability $|a_i|^2$ of returning to the ground state after measuring $P_0$.  
We now analyze the case that this measurement of $P_0$ reveals that we are not in the ground state; then the resulting state is equal to
\begin{eqnarray}
\label{transition}
&&\frac{1}{\sqrt{1-|a_i|^2}} (1-P_0) \phi_i \\ \nonumber &=& \frac{1}{\sqrt{1-|a_i|^2}}\phi_i-\frac{\overline a_i}{\sqrt{1-|a_i|^2}} \psi_0,
\end{eqnarray}
up to a phase.
Then, the probability that the subsequent measurement of $Q$ gives outcome $j$ is equal to
\begin{eqnarray}
\label{transition2}
&& |\langle \phi_j |  \frac{1}{\sqrt{1-|a_i|^2}}\phi_i-\frac{\overline a_i}{\sqrt{1-|a_i|^2}} \psi_0 \rangle|^2 \\ \nonumber
&=& \frac{|\delta_{i,j}-a_j \overline a_i|^2}{1-|a_i|^2},
\end{eqnarray}
and the resulting state is exactly equal to $\phi_j$.  Repeating this, we find that after every measurement of $Q$ with outcome $i$, the probability that the measurement of $P_0$ will return to the ground state is $|a_i|^2$, while if we do not return to the ground state the probability that the next measurement of $Q$ will give outcome $j$ is given by Eq.~(\ref{transition2}).

The transitions are then governed by a Markov chain with transition probabilities
\be
\label{Tii}
T_{i \leftarrow i}=(1-|a_i|^2)^2=1-2|a_i|^2+|a_i|^4,
\ee
\be
T_{j \leftarrow i, j \neq i}=|a_i|^2 |a_j|^2
\ee
\be
T_{0 \leftarrow i}=|a_i|^2,
\ee
where $T_{j \leftarrow i}$ is that, starting in state $\phi_i$, a measurement of $P_0$ reveals that one is not in the ground state, and then a measurement of $Q$ gives outcome $j$, while $T_{0 \leftarrow i}$ is the probability that, starting in state $\phi_i$, does leave on in the ground state.  The initial measurement of $Q$ at the start of the Markov chain gives state $\phi_i$ with probability $|a_i|^2$.  We would like to work out the expected number of measurements before the algorithm terminates back in the ground state.

To do this calculation, we re-interpret the probabilities, saying that with probability $2|a_i|^2$ the system makes some transition; half the time this transition leads to state $0$, while the other half the time the system makes a transition from state $i$ to some state $j$, and state $j$ is chosen with probability proportional to state $|a_j|^2$.  See the factor $-2|a_i|^2$ in Eq.~(\ref{Tii}) which we interpret as the negative of the probability of there being a transition;  note also that $j$ may equal $i$, so some of the ``transitions" do not actually lead to a change in state: this is the factor of $|a_i|^4$ in Eq.~(\ref{Tii}).  This re-interpretation only makes sense if $|a_i|^2\leq 1/2$, of course but it will allow us to compute the expected time exactly and as the expected time is analytic, this calculation will then work for all $|a_i|^2$.  The advantage of this re-interpretation is that if a transition occurs, and if the transition leads to a state other than $0$, then the probability that that state is equal to $j$ is proportional to $|a_j|^2$, which is the same probability distribution as the chain began in.

Starting in state $i$, the expected number of steps before a transition is $1/2|a_i|^2$.  Thus, averaging over states $i$ with initial probability distribution $|a_i|^2$, the expected number of steps before a transition is $k/2$, where $k$ is the number of possible outcomes of measurement $Q$.  Half of these transitions return to state $0$, while the other half do not, leading to some state $j\neq 0$ with probability distribution proportional to $|a_j|^2$.  Thus, the average number of steps before returning to the ground state is
\be
N_{steps}=k.
\ee
Note that both the number of phase estimations required and the number of measurements of $Q$ required are equal to $k$.

\subsubsection{Quadratic Speedup}
Suppose we wish to measure a projector $Q$.  
Define unitaries $U=2Q-1$ and $V=2P_0-1$.
Assume that $P_0$ has rank one.
Applying any sequence of projections $P_0$ or $Q$ to the ground state $\psi_0$ gives some state in the space spanned by the non-orthogonal basis vectors $\psi_0, Q \psi_0$.  Hence, we will work in this two-dimensional space for our analysis (the degenerate cases that $Q\psi_0=\psi_0$ or $Q\psi_0=0$ can be handled easily too and one can verify that the final result will be correct in these cases also).  In this subspace, in the non-degenerate case, we can write
\be
V=\begin{pmatrix} 1 & 0 \\ 0 & -1 \end{pmatrix},
\ee
\be
U=\begin{pmatrix} 2\cos^2(\theta)-1 & -2\cos(\theta)\sin(\theta) \\ -2\cos(\theta)\sin(\theta) & 2\sin^2(\theta)-1 \end{pmatrix},
\ee
where the angle $\theta$ is such that $\langle \psi_0 | Q | \psi_0 \rangle= \cos^2(\theta)$.
We emphasize that the above two equations are written in the {\it orthogonal} basis $\psi_0, Z (Q \psi_0-\cos^2(\theta) \psi_0)$, where $Z$ is
some normalization factor.
Thus, the unitary $UV$ restricted to the subspace has eigenvalues
\be
\exp(\pm 2 i \theta).
\ee

We now build a quantum circuit that adds an additional ancilla that controls both $U$ and $V$, hence controls whether or not $UV$ is applied.
We then implement phase estimation using this ancilla to determine the eigenvalues of $UV$.  We use the ground state $\psi_0$ as input to the phase estimation circuit.  Let $T_0$ be the time required to implement $UV$.  Using the ancilla to control $(UV)^n$, for various powers of $n$, one is able to measure the eigenvalue of $UV$ to accuracy $\epsilon$ in a time proportional to $T_0 \epsilon^{-1} \log(\epsilon^{-1})$; in this case, we take $n$ of order $\epsilon^{-1}$.

In fact, $UV$ has two distinct eigenvalues, and the phase estimation will return one of the two answers randomly.  However, since the eigenvalues differ only in sign, and the desired expectation value $\langle \psi_0 | Q | \psi_0 \rangle$ is equal to $\cos^2(\theta)$, both eigenvalues give the same answer for this expectation value.

After implementing this phase estimation to the desired accuracy, we then are left with some state in the given subspace.  If we then wish to recover the ground state (to recover some other observable), we can apply the recovery map above by alternating measurements of $Q$ and $P_0$ until the ground state is restored.

\subsubsection{Measuring $P_0$}
\label{MP0}
The above procedure relies on the ability to measure $P_0$.  The key is to define an operation that will only identify whether or not one is in the ground state, without revealing additional information.  If instead the measurement gives several bits of information on the energy of the state, the procedure does not work; intuitively, each $\phi_i$ is a coherent superposition of different energy eigenstates and making this more detailed measurement destroys this information.

We now describe how to do this; we call this procedure a ``coherent phase estimation".  We have the ability to perform a phase estimation on the Hamiltonian $H$ to determine the energy of a state.  This phase estimation is given by a quantum circuit, in which several control bits are initialized in a state $|0\rangle$.  Then, a Hadamard is applied to each control bit.  Then, the control bits are used to control the application of the Hamiltonian for a set period of time.  Finally, the Hadamard is re-applied, and the control bits are measured in the computational basis.  Then, classical post-processing is applied to extract the energy of the state.  This process can be performed in serial (using only one control bit) or in parallel; the parallel approach uses more control bits but reduces the depth of the circuit. 
The measurement of the energy is not deterministic; however, one can employ a larger number of control bits to increase the accuracy. 

To measure $P_0$,
we use this phase estimation in its parallel form, but do not measure the outcomes of the control bits.  We initialize an additional ``outcome" bit to $|0\rangle$.  We then use a unitary quantum circuit to implement the classical postprocessing done to determine the energy, storing that energy value coherently.  Then a unitary quantum circuit determines if that energy is near the ground state value (which we have determined in advance by a phase estimation before doing any measurements), and if so, it flips the value of the outcome bit.  Finally, we uncompute, reversing the classical postprocessing and phase estimation steps, and then at the end we measure the outcome bit.  This procedure is essentially that in Ref.~\onlinecite{PV}.
While this procedure does not implement a projective measurement, but rather implements a POVM, if sufficient numbers of control bits are used, the measurement can be arbitrarily close to the desired projective measurement.

Performing this measurement of $P_0$ does require additional qubits.  The number of qubits required scales proportional to the number of bits of accuracy desired; however, since this accuracy is of order $\Delta$, if $\Delta$ is only polynomially small in system size then this requires only logarithmically many extra bits.  Additionally, the more accurately we implement a projective measurement rather than a POVM, the more control bits required, but this overhead also only scales logarithmically.

An alternate method of implementing $P_0$ is to note that we can easily implement the projector onto the ground state of the Hamiltonian at the start of the annealing process if our initial Hamiltonian is of a sufficiently simple form, such as a free fermion Hamiltonian or a Hamiltonian with a product ground state.  For a general free fermion Hamiltonian, we can use Givens rotations to rotate to the case that the ground state simply has some spin-orbitals occupied and some empty; we then measure whether we have the desired pattern or not.  Call this projector $P_0^{ff}$.  Then, if $U_{adiab}$ is the unitary which describes the approximately adiabatic evolution from the free Fermion Hamiltonian onto the desired final Hamiltonian, we can approximate $P_0 \approx U_{adiab} P_0^{ff} U_{adiab}^\dagger$.  The accuracy of this approximation depends upon the annealing path; see the discussion in ~\ref{Sec:anneal}, which
shows that
we can use this strategy to achieve a (near) quadratic speedup.  The improved annealing paths discussed there are most useful for the particular approach here where we need very high accuracy in the annealing; in all other applications in the paper, we need only moderate accuracy in the annealing, sufficient to get a large overlap with the ground state so that phase estimation can then project onto the ground state with reasonably large probability.  We therefore

\section{Conclusions}

In this paper we have presented a comprehensive strategy for using quantum computers to solve models of strongly correlated electrons, using the Hubbard model as a prototypical example. Similar strategies can be used for generalizations of the Hubbard model and other discrete quantum lattice models, including, but not limited to, the $t$-$J$ model, frustrated magnets, or bosonic models coupled to (static) gauge fields, which all suffer from negative sign problems in quantum Monte Carlo simulations on classical computers. We go beyond previous papers \cite{Abrams:1997ha,Abrams:1999jv,Ortiz2001,Somma2002}  that discussed simulations of the Hubbard model on quantum computers by giving complete details of all steps needed to learn about the properties of the Hubbard model.

 In particular we presented a strategy to prepare trial ground states starting from various mean-field states or RVB states on decoupled plaquettes. While finding the ground state of quantum lattice models is QMA-hard\cite{Kitaev02,qma1d,qmabh} this is a statement about the worst case. We expect that, as experience has shown, many experimentally relevant models have ground states with either no long-range order (so-called quantum liquids) or with broken symmetries that are qualitatively well described by mean-field theories. In fact, if a model has such peculiar properties that it is hard to find its ground state on a quantum computer, then also a material described by that model will have a hard time reaching its ground state. Quick low-temperature thermalization in experiments is thus an indication that the ground state of the model describing its properties should also be easy to prepare on a quantum computer. The opposite is the case, e.g. for (classical) spin glasses, where finding the ground state is NP-hard but also spin glass materials never thermalize nor reach the ground state.
 
We have furthermore presented an efficient and deterministic quantum algorithm to prepare arbitrary Slater determinants as initial state, which scales better than the algorithms of Refs. \onlinecite{Ortiz2001,Somma2002}. This can be used to prepare ground states of various candidate mean-field Hamiltonians, from which an adiabatic evolution to the ground state of the Hubbard model can be attempted. 

To implement time evolution under the Hubbard and mean-field Hamiltonians we have given explicit quantum circuits for all terms, and discussed the size of the Trotter time step required to achieve sufficiently small errors. We discuss how time evolution under the individual terms in the Hubbard model can be efficiently parallelized, ultimately requiring  $\mathcal{O}(N)$ qubits and gates with only $\mathcal{O}(\log N)$ parallel circuit depth for one time step, which allows efficient simulations of very large systems. We gain additional constant factors over previous approaches, by optimizing the phase estimation algorithm to reduce the required number of (expensive) rotation gates by a factor of four, assuming that our gate set consists of one- or two-qubit Clifford operations and arbitrary single qubit rotations. We furthermore propose to use a larger Trotter time step for the adiabatic state preparation (whose time scale is controlled by the inverse gap squared), to prepare a very good (but not perfect) guess for the ground state, and then refine this to the exact ground state by doing only the final quantum phase estimation (whose time scale is shorter since it is controlled by the inverse gap) with a small time step.

We have finally discussed approaches to obtain a quadratic speedup in measurements by proposing two non-destructive measurement strategies, one based on the Hellman-Feynman theorem, and another based on recovering the ground state after a (destructive) measurement of a single qubit. We also introduce a new approach of measuring dynamic structure factors and spectral functions directly in frequency space, using ideas similar to angle-resolved photoemission experiments.

Estimating the gate counts required to simulate the Hubbard model, we find that even on lattices with more than $N\approx1000$ sites -- which should be large enough to learn most interesting properties -- can be simulated on small scale quantum computes using only a few thousand qubits and a parallel circuit depth of about one million gates. This number is based on our estimate of a few thousand time steps with a circuit depth of not more than a few hundred  gates each. This means that even with logical gate times of the order of 1$\mu s$, the ground state of the Hubbard model can be prepared within a few seconds. Small scale quantum computers will thus be a very powerful tool for the investigation of many problems in the field of strongly correlated electron models that are currently out of reach of any classical algorithm.

\acknowledgements

We acknowledge useful discussions with Bela Bauer. This work was supported by Microsoft Research, by the Swiss National Science Foundation through the National Competence Center in Research NCCR QSIT, and the ERC Advanced Grant SIMCOFE. The authors acknowledge hospitality of the Aspen Center for Physics, supported by NSF grant \# 1066293.


\bibliographystyle{apsrev4-1}
\bibliography{paper}

\appendix
\section{Error bounds for time-dependent Trotter Formulas}

Simulating the dynamics of time--dependent Hamiltonians on quantum computers is a more subtle issue than simulating time--independent dynamics and results proven for the time--independent case do not necessarily transfer over.  This issue is significant here because of issues surrounding simulating adiabatic state preparation.  Bounds for the error in high--order Trotter formulas are known, however, the scaling of these bounds with the number of terms in the Hamiltonian is known to be loose.  Since the Hamiltonians needed for our purposes have a large number of terms, tighter estimates of the error scaling may be important for determining the cost of performing adiabatic state preparation for Fermionic systems.  

It may be tempting to think that we can neglect the errors in the Trotter formula that are incurred by approximating $H(t)$ by a piecewise constant Hamiltonian
 because the time derivatives of the Hamiltonian are small for a slow adiabatic evolution, but this is not necessarily true because adiabatic theorems only require that the derivatives of the Hamiltonian are small relative to an appropriate power of the minimum eigenvalue gap.  This means that in some circumstances adiabatic evolution may actually involve a rapid passage.   Furthermore, since we are interested in high accuracy state preparation even though such errors may be small they will not necessarily be negligible compared to our target error tolerance. 

In this section we analyze these errors in more detail.
In particular, we will examine the error in the second order order Trotter--Suzuki formula 
\begin{equation}
\mathcal{T}e^{-i\int_0^1 H(u) \mathrm{d}u} \approx \prod_{j=1}^m e^{-iH_j(1/2)/2}\prod_{j=m}^1 e^{-iH_j(1/2)/2},
\end{equation}
where we approximate the evolution from time $0$ to $1$ by a single second-order Trotter-Suzuki step.
The bounds that we present are a generalization of those in Ref.~\onlinecite{Wecker14a} to the time--dependent case.  Bounds for the Trotter formula are given in Ref \onlinecite{HdR90}.

Our main result is that
\begin{eqnarray}
\label{appendixmainresult}
&&\left\| {\cal T} e^{-i \int_0^1 H(u) {\mathrm d}u} - \prod_{j=1}^m e^{-iH_j(1/2)/2}\prod_{j=m}^1 e^{-iH_j(1/2)/2} \right\| \nonumber \\ \nonumber
&\leq & \frac{1}{24} {\rm max}_s \| H''(s) \| + \frac{1}{12}\| [H'(1/2),H(1/2)] \| \\ \nonumber
&& +\sum_{j=1}^m \left\|\left[\left[H_j(1/2),\sum_{k>j} H_k(1/2)\right],H_j(1/2)\right] \right\| \\
&&+ \left\|\left[\left[\sum_{k>j} H_k(1/2),H_j(1/2)\right],\sum_{\ell>j} H_\ell(1/2)\right] \right\|.
\end{eqnarray}
This result reduces to the bound of~\onlinecite{Wecker14a} and agrees with the asymptotic error scaling predicted in ~\onlinecite{Babbush15}, up to a use of the triangle inequality and a small multiplicative factor, if $H$ is time-independent.

We prove the bound in two steps.  First, we will show that 
\begin{eqnarray}
\label{deriverror}
&&\| {\cal T} e^{-i \int_0^1 H(u) {\mathrm d}u} - e^{-i H(1/2)} \|
\\ \nonumber
&&\quad\leq \frac{1}{24} {\rm max}_s \| H''(s) \| + \frac{1}{12}\| [H'(1/2),H(1/2)] \|.
\end{eqnarray}
Then, we use the bound of Ref.~\onlinecite{Wecker14a} that
\begin{eqnarray}
&&\Bigr\| e^{-iH(1/2)} - \prod_{j=1}^m e^{-iH_j(1/2)/2}\prod_{j=m}^1 e^{-iH_j(1/2)/2} \Bigr\| \nonumber
\\
& \leq &\sum_{j=1}^m \left\|\left[\left[H_j(1/2),\sum_{k>j} H_k(1/2)\right],H_j(1/2)\right] \right\| \\ \nonumber
&&+ \left\|\left[\left[\sum_{k>j} H_k(1/2),H_j(1/2)\right],\sum_{\ell>j} H_\ell(1/2)\right] \right\|.
\end{eqnarray}
 Eq.~(\ref{appendixmainresult}) then follows from the triangle inequality.  We remark that in Ref.~\onlinecite{Wecker14a}, a typographical error led to $\|[[A,H(x)],H(x)] \| \leq \|[[A,B],A]\|+\|[[A,B],B]\|$ being replaced by $\|[[A,H(x)],H(x)] \| \leq \|[[A,B],A]]+[[A,B],B]\|$ in the language of that paper in appendix B (see text above Eq.~(B3) of that paper).  This pair of missing $\|$ symbols propagated to later works although it does not affect any of the scaling results cited in later papers.  In the above equation, we have corrected this error, so that the right hand-side is a sum of two distinct norms for each $j$, rather than a norm of a sum.

To show Eq.~(\ref{deriverror}),
we write $H',H'',...$ to denote the first, second, ... derivatives of $H$ with respect to time.
By Taylor's theorem, $H(u)=H(1/2)+(u-1/2) H'(1/2) + \int_{1/2}^u (u-s) H''(s) {\rm d}s$.
Hence $\mathcal{T}e^{-i\int_0^1 H(u) \mathrm{d}u}$ is
\begin{eqnarray}
{\cal T} e^{ -i\int_0^1 \Bigl( H(1/2)+(u-1/2) H'(1/2) + \int_{1/2}^u H''(s) (u-s) {\rm d}s \Bigr){\mathrm d}u},\nonumber
\end{eqnarray}
and thus
\begin{eqnarray}
\label{firstbound}
&& \Bigr\| {\cal T} e^{-i \int_0^1 H(u) {\mathrm d}u} -{\cal T} e^{ -i\int_0^1 \Bigl( H(1/2)+(u-1/2) H'(1/2) \Bigr)\mathrm{d}u} \Bigr\| \nonumber \\ \nonumber
&&\qquad \leq  \int_0^1 \left\|  \int_{1/2}^u H''(s) (u-s) {\rm d}s \right\| {\mathrm d}u \\
&& \qquad \leq  \frac{1}{24} {\rm max}_s \| H''(s) \|.
\end{eqnarray}

We next bound the difference
$$\left\| {\cal T} e^{-i \int_0^1 \Bigl( H(1/2)+(u-1/2) H'(1/2) \Bigr)\mathrm{d} u} - e^{-i H(1/2)} \right\|.$$
Transforming to the interaction representation gives
\begin{eqnarray}
&&{\cal T} e^{-i \int_0^1 \Bigl( H(1/2)+(u-1/2) H'(1/2) \Bigr)\mathrm{d}u}  \nonumber\\
&&\qquad= e^{-i (1/2) H(1/2)} \Bigl( {\cal T} e^{-i \int_0^1 (u-1/2) H'_u(1/2) {\mathrm d}u }\nonumber
\Bigr)
\\ 
&&\qquad~ \times
e^{-i (1/2) H(1/2)},\nonumber
\end{eqnarray}
where we define
\begin{eqnarray} \nonumber
&& H'_u(1/2)
\equiv
e^{-i (1/2-u) H(1/2)} H'(1/2) e^{i (1/2-u) H(1/2)}.
\end{eqnarray}
So by a unitary rotation
\begin{eqnarray}
&& \| {\cal T} e^{-i \int_0^1 \Bigl( H(1/2)+(u-1/2) H'(1/2) \Bigr)\mathrm{d}u} -e^{-i H(1/2)} \|\label{secondboundpart1} \\
&&\quad  =  \| {\cal T} e^{-i \int_0^1 (u-1/2)H'_u(1/2) {\mathrm d}u } -1 \|\nonumber\\
&&\quad  =  \| {\cal T} e^{-i \int_0^1 (u-1/2)H'_u(1/2) {\mathrm d}u } -{\cal T} e^{-i \int_0^1 (u-1/2)H'(1/2) {\mathrm d}u } \|.\nonumber
\end{eqnarray}
Note that $\|H'_u(1/2)-H'(1/2)\|$ can be written as
\begin{eqnarray}
\left\|\int_0^1\! \partial_x\! \left(e^{-i (\frac 1 2-u)(1-x) H(\frac 1 2)} H'(\frac 1 2) e^{i (1/2-u)(1-x) H(\frac 1 2)}\right)\! \mathrm{d}x\right\|\nonumber
\end{eqnarray}
which can be used to show that
 $\| H'_u(1/2) -  H'(1/2) \| \leq |u-1/2| \| [H'(1/2),H(1/2)] \|$, and so
\begin{eqnarray}
\label{secondbound}
 && \| {\cal T} e^{-i \int_0^1 (u-1/2) H'_u(1/2) {\mathrm d}u }-{\cal T} e^{-i \int_0^1 (u-1/2)H'(1/2) {\mathrm d}u } \|\nonumber
\\ \nonumber
&&\quad \leq 
\int_0^{1} |u-1/2|^2 \| [H'(1/2),H(1/2)] \|\mathrm{d}u \\ 
&&\quad= \frac{1}{12}\| [H'(1/2),H(1/2)] \|.
\end{eqnarray}

 Eq.~(\ref{deriverror}) then follows from Eqs.~(\ref{firstbound},\ref{secondboundpart1},\ref{secondbound}) and the triangle inequality.

\end{document}